\documentclass{WileyMSP-template}

\usepackage{amsmath} 
\usepackage{xcolor} 
\usepackage{upgreek} 
\let\Gamma\varGamma 
\let\Delta\varDelta 
\usepackage[labelfont=bf]{caption} 
\usepackage{textcomp} 
\usepackage{lineno} 
\usepackage{booktabs,caption}

\begin{document}

\pagestyle{fancy}

\title{Large Bidirectional Refractive Index Change in Silicon-rich Nitride via Visible Light Trimming}

\maketitle

\author{Dmitrii Belogolovskii*,}
\author{Md Masudur Rahman,}
\author{Karl Johnson,}
\author{Vladimir Fedorov,}
\author{Andrew Grieco,}
\author{Nikola Alic,}
\author{Abdoulaye Ndao,}
\author{Paul K. L. Yu,}
\author{and Yeshaiahu Fainman}

\begin{affiliations}
Dr. D. Belogolovskii, M. M. Rahman, K. Johnson, V. Fedorov, Dr. A. Grieco, Dr. N. Alic, Prof. A. Ndao, Prof. P. K. L. Yu, Prof. Y. Fainman\\
University of California, San Diego, 9500 Gilman Drive, La Jolla, CA 92093, USA\\
dbelogol@ucsd.edu

\end{affiliations}

\keywords{Silicon nitride; integrated photonics; optical trimming; laser annealing; tunable optical material}

\begin{abstract}

Phase-sensitive integrated photonic devices are highly susceptible to minor manufacturing deviations, resulting in significant performance inconsistencies. This variability has limited the scalability and widespread adoption of these devices. Here, a major advancement is achieved through continuous-wave (CW) visible light (405 nm and 520 nm) trimming of plasma-enhanced chemical vapor deposition (PECVD) silicon-rich nitride (SRN) waveguides. The demonstrated method achieves precise, bidirectional refractive index tuning with a single laser source in CMOS-compatible SRN samples with refractive indices of 2.4 and 2.9 (measured at 1550 nm). By utilizing a cost-effective setup for real-time resonance tracking in micro-ring resonators, the resonant wavelength shifts as fine as 10 pm are attained. Additionally, a record red shift of 49.1 nm and a substantial blue shift of 10.6 nm are demonstrated, corresponding to refractive index changes of approximately 0.11 and $-2\times10^{-2}$. The blue and red shifts are both conclusively attributed to thermal annealing. These results highlight SRN’s exceptional capability for permanent optical tuning, establishing a foundation for stable, precisely controlled performance in phase-sensitive integrated photonic devices.  

\end{abstract}

\section{Introduction}

Silicon-rich nitride (SRN) grown using plasma-enhanced chemical vapor deposition (PECVD) has numerous qualities that are outstanding for a versatile complementary metal-oxide semiconductor (CMOS) compatible integrated photonics platform. As the Si content (and with it – the refractive index) of SRN films vary with the choice of PECVD deposition and post-processing parameters, films can be grown with a refractive index (as measured at 1550 nm) ranging from 2.0 (lower Si content: Si\textsubscript{3}N\textsubscript{4}) to 3.2 (higher Si content: Si\textsubscript{7}N\textsubscript{3}) [1,2]. Furthermore, other linear and nonlinear optical properties of SRN are also altered remarkably with the change of silicon content, including linear optical losses and transparency window [2], nonlinear optical losses [1,3], thermo-optic coefficient [4], and third-order susceptibility ($\chi_3$) [1,5]. Consequently, control of these properties enables SRN films to be optimized for various applications. For example, a wide variety of nonlinear effects have been demonstrated in SRN devices, such as the DC Kerr effect [6], nonlinear Kerr switching [7], second harmonic generation [8], intermodal frequency generation [9], four-wave mixing [5], and supercontinuum generation [10]. Additionally, SRN’s relatively high thermo-optic coefficient (especially for higher Si-content films) enables implementation of efficient thermally driven phase shifters and switches [11,12]. Finally, SRN is substantially more compatible with traditional CMOS mass-manufacturing process than other materials, such as low pressure chemical vapor deposition (LPCVD) Si\textsubscript{x}N\textsubscript{y}, since the relatively low temperature of PECVD SRN deposition (350 °C) does not exceed the thermal budget of front-end-of-the-line semiconductor processes [2].
\\~\\
On the other hand, fabrication variability continues to represent a persistent challenge in integrated photonics, including SRN integrated platform. In particular, phase-sensitive devices, such as micro-ring resonators (MRRs), are highly sensitive to variations in the fabrication process even on the scale of 1 nm, which can lead to significant performance inconsistencies, ultimately precluding these devices from commercial and widespread use [13,14]. These random variations typically require optical devices to be biased to operating specifications using power-demanding heaters [15]. Alternatively, optical trimming, which is fine tuning of the optical properties of photonic devices post-fabrication, has been employed to compensate for such variations. However, most presented trimming methods either induce unidirectional refractive index change only, or suffer from strong relaxation, or require CMOS-incompatible materials as cladding, limiting their effectiveness.
\\~\\
It has been well-known for nearly half a century that optical materials exposure to high-intensity visible or ultra violet (UV) light can alter their optical properties [16]. As such, some of the first approaches to trimming integrated photonic devices used UV exposure to silane-based waveguide cladding — and soon after, Si\textsubscript{x}N\textsubscript{y} waveguide cores — to substantially shift the center wavelengths of filters and ring resonators [17,18]. In the time since, a wide variety of methods and optical materials have been proposed to trim optical properties of integrated components, including further developments of continuous wave (CW) UV and visible light exposure [19-22], pulsed laser irradiation [23], electron beam exposure [24,25], and localized annealing of regions exposed to ion implantation [26-31]. These methods can introduce waveguide effective index changes on the order of $10^{-3}$ to $10^{-2}$, which is sufficient for many applications [26,30]. However, several of these methods require exotic non-CMOS-compatible materials to sensitize the optical structure to the applied stimulus, such as chalcogenides, phase change materials, tunable polymers, and liquid crystals [17,19,21,32]. Other methods require trimming to be performed on waveguides prior to the application of the final cladding, necessitating the use of temporary cladding and adding complexity to the trimming process [33]. Recently, simple CMOS-compatible post-fabrication trimming of structures has been demonstrated using localized annealing from on-chip heaters included in the chip design [26-31]. While this method is simple and effective, it requires Ge implantation of the waveguide and is capable only of decreasing the refractive index of the optical material. In fact, most of those mentioned above techniques are not capable of post-fabrication trimming in both directions.
\\~\\
Exploring the feasibility of trimming SRN via laser annealing is of significant interest. It is widely recognized that rapid thermal annealing (RTA) of SRN at high temperatures (typically \textgreater 600 °C) leads to the dissociation of Si-H bonds, significant hydrogen (H) desorption, reduced nitrogen (N) content, and an increased concentration of Si-Si bonds [34-36]. Conversely, studies have shown that RTA at lower temperatures (\textless 600 °C) causes bond redistribution (such as Si-Si, Si-N, Si-H, N-H), leading to an increase in Si-H and Si-N bond concentration at the expense of N-H and Si-Si bond concentration [34,35]. Such structural changes should inevitably result in permanent refractive index change necessary for trimming. At the same time, a relatively low thermal conductivity of PECVD SRN thin film was reported, in the range of 0.45 to 4.5 W m\textsuperscript{-1} K\textsuperscript{-1}[37], much lower then that of c-Si, which is 148 W m\textsuperscript{-1} K\textsuperscript{-1}[38]. This suggests that SRN can efficiently localize heat, which is favorable for localized laser annealing where high temperatures are required. When a thin film absorbs laser radiation, the resulting temperature increase can be approximately estimated using the equation $T_{max} = T_{0}+P/(w_{0} k \sqrt{\pi})$, where $T_{max}$ is the maximum temperature, $T_{0}$ is the room temperature, $w_{0}$ is the mode diameter of the Gaussian beam, $k$ is the thermal conductivity, and $P$ is the optical power [39]. Assuming that SRN's thermal conductivity is roughly in the middle, for example, $k$ = 1.7 W m\textsuperscript{-1} K\textsuperscript{-1}, and $P$ = 40 mW, $w_{0}$ = 10 $\mu$m, we estimate $T_{max}$ = 1350 °C, large enough to activate both mechanisms that are responsible for thermal annealing [34-36], making SRN an excellent candidate for trimming studies.
\\~\\
In this manuscript, we present extensive and novel results on visible CW radiation (405 nm and 520 nm) induced bidirectional trimming of PECVD SRN with refractive indices of 2.4 and 2.9 (at 1550 nm). Refractive index tuning is accomplished in a simple and cost-effective setup that allows rapid fine trimming via real-time position tracking of a resonance in an MRR. Remarkably, both 405 nm and 520 nm radiation can induce blue and red wavelength resonance shifts in an SRN MRR. The latter implies that a refractive index of SRN can be both decreased and increased by using a single laser-source, which is undoubtedly a highly advantageous feature for the trimming of SRN devices. Furthermore, the blue and red shifts can be easily separated in a fully controllable way simply by changing the trimming laser power or the exposure duration. Moreover, a strong blue shift of 10.6 nm and a largest ever reported red (to date) shift of 49.1 nm are achieved in trimmed SRN MRRs. In addition, the refractive index can be decreased by $-2\times10^{-2}$ and increased by a large value of 0.11. We demonstrate that both blue and red shifts originate from thermal annealing, with blue shifts occurring at lower annealing temperatures (400 °C – 500 °C) and red shifts emerging at temperatures above 500 °C – 600 °C. Intriguingly, the magnitude and direction of the shifts vary significantly with the refractive index of SRN, which makes the material an excellent choice for coating. Additionally, it is established that the presence of both red and blue shifts mitigates the resonance backshifting by pushing the resonance in opposite directions. Finally, we demonstrate the effectiveness of the novel technique employing it for trimming of an SRN-based optical demultiplexer with MRRs to an accuracy of approximately 10 pm. Overall, our results suggest that SRN is a highly reconfigurable CMOS-compatible platform, which offers a pathway toward significantly improving the precision and manufacturability of SRN (or Si if SRN is used as coating) photonic devices, particularly for phase-sensitive applications where sub-nanometer tuning accuracy is required.

\section{Results} 
\subsection{Experimental Setup}

\begin{figure}[htbp]
\includegraphics[width=1\linewidth]{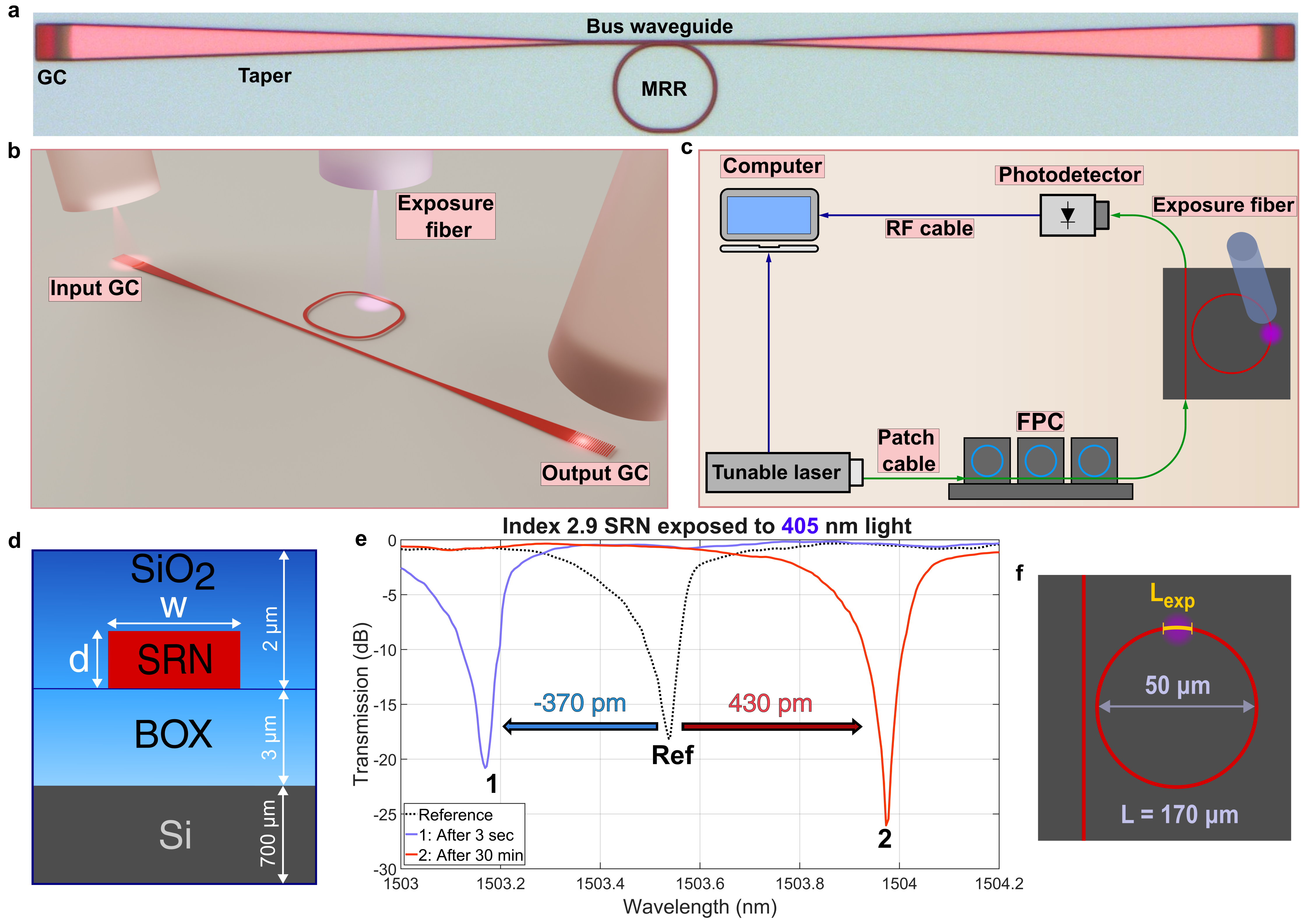}
\caption{(a) Microscope image of a device to be trimmed showing an MRR, bus waveguide, grating couplers (GC), tapers. (b) 3D illustration of the configuration of 3 fibers used for trimming and real-time measurement. Laser light couples in and out via grating couplers. The exposure fiber is centered over a section of an MRR. (c) Block-diagram of the setup showing the equipment used in experiments. Here FPC is fiber polarization controllers used to set transverse electric (TE) polarization. Blue arrows denote RF cables, while the green ones – fiber patch cables. (d) Cross-section of a waveguide used in experiments. Here BOX is buried oxide. The parameters $w$ and $d$ are 600 nm and 385 nm for the index 2.4 SRN, respectively, while they are 450 nm and 340 nm for the index 2.9 SRN, respectively. (e) Transmission spectra of the probing signal in SRN MRRs after exposure to 405 nm light. The dotted black curve indicates the initial position of resonance (reference). The numbers (and "Ref") next to the resonances indicate the order in which the spectra were measured (starting with "Ref"). The labels describe exposure time. (f) Image of MRR showing how its section is exposed (violet circle), MRR diameter is 50 $\mu$m, and ring length ($L$) is 170 $\mu$m. The length of one exposed section is $L_{exp}$.
\label{fig1}}
\end{figure}

To investigate the effects of visible light exposure on SRN waveguides, we deposited SRN films with refractive indices of 2.4 and 2.9 (measured at 1550 nm) using PECVD. For the waveguide with a refractive index of 2.4, the width and thickness were 600 nm and 385 nm, respectively; for the waveguide with an index of 2.9, they were 450 nm and 340 nm. The fabricated MRRs had a radius of 25 $\mu$m, with a coupler length of 6.5 $\mu$m, yielding a total length of approximately 170 $\mu$m. The gap between a bus waveguide and a coupler was 400 nm. \textbf{Figure 1}a demonstrates a microscope image of the fabricated device used for trimming experiments.
\\~\\
Figure 1b and 1c demonstrate the setup used for the trimming experiments. The goal was to achieve a strong MRR resonance shift and fine-tuning in real time as the visible laser exposure was applied, while maintaining a simple and cost-effective setup. For this, we used inexpensive fiber-coupled multimode (MM) 405 nm and 520 nm laser diodes (Laser Tree LT-FCLD-M405075 and LT-FCLD-M520065) coupled to SMF-28 patch cables. The bare fiber tip, cleaned and cleaved, was angled at 10 degrees relative to the sample to minimize back reflections. The fiber was attached to the grating coupler stage which had six degrees of freedom, enabling precise alignment, with an accuracy of 500 nm (more details on how high accuracy alignment was achieved can be found in Supporting Information, section S1). The trimming resolution (that is the minimum feature size that can be exposed) in this setup was 10 $\mu$m, primarily limited by the beam size of the trimming light. While the SMF-28 patch cable was not optimized for 405 nm or 520 nm light, we found it suitable, with transmission losses of only 0.7 dB. No degradation in cable or laser performance was observed, even at maximum power of 16.4 dBm used in the experiments. It is worth noting that one way to improve the trimming resolution is to use a single-mode (SM) fiber specifically designed for 405 nm, such as the S405-XP, which has an MFD of 3.3 µm and is sufficient for many applications. Additionally, using a lensed 405 nm SM fiber could further reduce the size of the trimming beam.
\\~\\
Figure 1c demonstrates a block-diagram of the setup used in the experiments. To monitor the trimming-induced resonance shift in real time, light from an Agilent 81980A tunable continuous-wave laser (1465 nm – 1575 nm) was coupled into the SRN samples (cross-section is shown in Figure 1d) using grating couplers. Polarization controllers were used to set polarization to transverse electromagnetic (TE) mode, and the output light was measured by a photodetector. 
\\~\\
Figure 1e exemplifies the measured transmission spectra of the probing signal in SRN MRRs when a section of an MRR was exposed to 405 nm (as illustrated in Figure 1f; also see Supporting Information, section S2, for more results). The results confirm that it is indeed possible to achieve both red and blue resonance shifts using a single laser source. A blue shift was observed within a few seconds of exposure while the red shift took up to 30 minutes to saturate. These results demonstrate the ability to clearly separate blue and red shifts by varying the duration of exposure, a crucial property for trimming applications. 

\subsection{Dynamics and Power Dependence of SRN Refractive Index Change} 

This section systematically analyzes how key parameters — such as a refractive index of ($n_{srn}$), exposure wavelength ($\lambda_{exp}$), optical power of exposure ($P_{exp}$), exposure duration ($t_{exp}$) — affect the trimming process.  Specifically, we exposed SRN MRRs with refractive indices $n_{srn}$ of 2.4 and 2.9 to visible radiation wavelengths of 405 nm (violet) and 520 nm (green). 
\\~\\
The presence of both blue and red resonance shifts in MRRs demonstrates that the effective index of SRN waveguides $n_{eff}$ can be both decreased and increased using either 405 nm or 520 nm lasers. The refractive index change is proportional to the resonance wavelength shift, as shown in Equation (1-3) [39, 40]: 

\begin{equation}
\Delta n_{\text{eff}} = n_g \frac{\Delta \lambda_{\text{res}}}{\lambda_{\text{res}}} \frac{L}{L_{\text{exp}}}
\label{eq:1}
\end{equation}

\begin{equation}
n_g = \frac{\lambda_{\text{res}}^2}{\text{FSR} \cdot L}
\label{eq:2}
\end{equation}

\begin{equation}
L_{\text{exp}} = \frac{\Delta \lambda_{\text{sec}}}{\Delta \lambda_{\text{ring}}} L
\label{eq:3}
\end{equation}

Here $\Delta n_{eff}$ is the change of the effective index $n_{eff}$ in an SRN waveguide, $n_{g}$ is the group index, $\lambda_{res}$ is the resonance wavelength of an MRR, $\Delta \lambda_{eff}$ is the resonance wavelength shift in an MRR, FSR is a free spectral range of an MRR, $\Delta \lambda_{ring}$ is the resonance wavelength shift when the whole MRR is exposed, $\Delta \lambda_{sec}$ is the resonance wavelength shift when only one section of the MRR is exposed, $L_{exp}$ is the length of the single exposed section (see Figure 1f), $L$ is the length of the MRR. Additionally, to estimate a refractive index change $\Delta n_{srn}$, we performed simulations in Lumerical to calculate what refractive index change $\Delta n_{srn}$ is required to cause the corresponding effective index change $\Delta n_{eff}$. Based on Equation (1-3), and the resonance shift from MRRs transmission spectra (like in Figure 1e, also see Figure S2), we estimated the refractive index changes. It is worth noting that $L_{exp}$ of 10 $\mu$m was estimated experimentally (see next section) and used in Equation (1-3). 
\\~\\
\begin{figure}[htbp]
\includegraphics[width=1\linewidth]{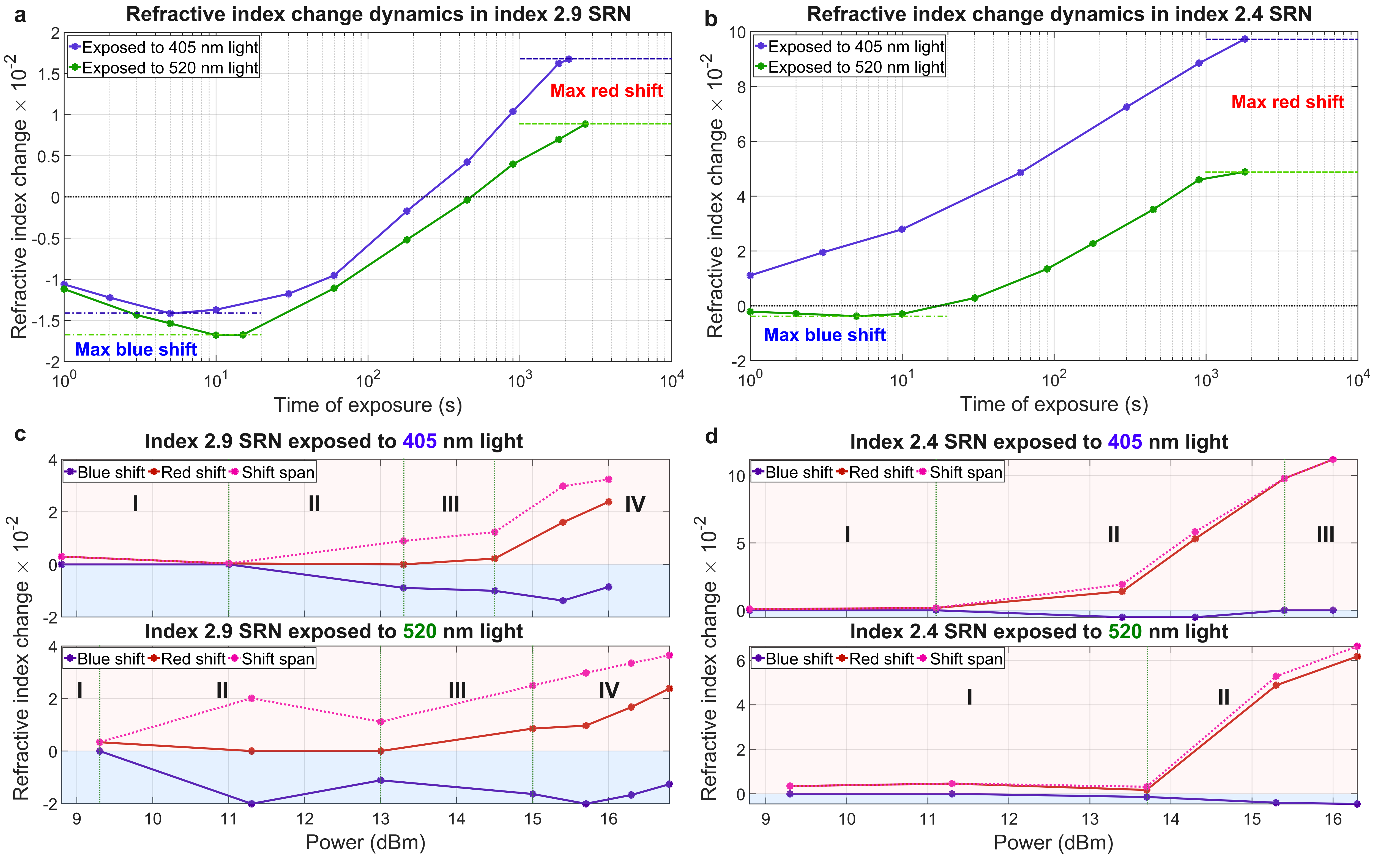}
\caption{Dynamics of resonance shift in an SRN MRR when (a) $n_{srn}$ = 2.9, and (b) $n_{srn}$ = 2.4. The exposure power $P_{exp} \approx 15.5\; dBm$; the violet curve represents exposure to 405 nm light, and the green – to 520 nm light; the dotted black line indicates no refractive index change. Refractive index change of SRN as a function of exposure power when (c) $n_{srn}$ = 2.9 and (d) $n_{srn}$ = 2.4; the upper plots represent the case when the exposure was done with the 405 nm laser, and the bottom ones – with the 520 nm laser; red and blue areas mark positive and negative refractive index changes, respectively.
\label{fig2}}
\end{figure}

First, we investigate the SRN refractive index change dynamics measured in MRRs exposed to 405 nm and 520 nm light (\textbf{Figure 2}a, 2b). Only one section of an MRR was exposed in each experiment, as shown in Figure 1f. In Figure 2a, we observe the refractive index change in an SRN MRR with $n_{srn}$ = 2.9 as a function of exposure time for 405 nm and 520 nm light. Focusing on the 405 nm case, the blue shift (due to refractive index decrease) saturates in approximately 5 seconds, while the red shift (due to refractive index increase) fully compensates for the blue shift in about 200 seconds when exposed to 405 nm light. In contrast, the red shift takes approximately 30 minutes to saturate. Meanwhile, both the blue and red shifts induced by the 520 nm laser follow a similar pattern, though they take slightly longer to saturate compared to the 405 nm laser. Also, stronger blue shift can be observed when induced by 520 nm laser, while strong red shift can be observed when induced by 405 nm laser (for the same power of exposure).
\\~\\
Similarly, Figure 2b illustrates the refractive index change in an SRN MRR with $n_{srn}$ = 2.4 as a function of exposure time for 405 nm and 520 nm light. This time, the difference in dynamics between the two lasers is more evident. Notably, no blue shift is observed within the first second of exposure to 405 nm light, unlike with the 520 nm laser. Additionally, a stronger red shift can be seen when an SRN MRR is exposed to 405 nm light.
\\~\\
The complex behavior of SRN trimming originates from the thermal annealing mechanism driving the refractive index change. We found that lower temperature annealing (400 °C – 500 °C) leads to a decrease in refractive index, while annealing at temperatures above 500 °C – 600 °C results in an increase. Notably, the refractive index reduction is significantly smaller in SRN with an index of 2.4 compared to SRN with an index of 2.9, and that agrees well with the results shown in Figure 2c and 2d, where it is evident that the blue shift is noticeably weaker in index 2.4 SRN. Greater details and discussion are provided in the Supporting Information, section S3. Also, in Supporting Information, section S6, we demonstrate that neither PECVD SiO\textsubscript{2} nor residual HSQ contribute to the permanent shifts observed in SRN MRRs.
\\~\\
It was also of great interest to investigate how exposure power affects direction of the shifts (or refractive index change) and their magnitude. Figure 2c and Figure 2d show the refractive index change in SRN measured in MRRs exposed to 405 nm and 520 nm light as a function of exposure power.
\\~\\
Based on Figure 2c, overall, we indicate 4 different sections with distinct dynamics (I – IV). In section I, at the lowest exposure powers, we observe only weak red shift, and no blue shift. As power increases, in section II, no red shift can be observed, although noticeable blue shift can be visible. It takes 10 min – 30 min for the blue shift to saturate. As the exposure power continues to increase, in section III, aside from a strong blue shift, the onset of the red shift can be seen. The red shift takes over only after the blue shift saturates (after about 5 min – 15 min), and it takes the red shift about 30 min to saturate. Therefore, it is possible to separate the directions of the shifts by time of exposure, although the red shift is much weaker since both red and blue shifts partially compensate each other. Finally, section IV reveals that both strong blue and red shifts are possible. In noticeable contrast to section III, the red shift is much stronger, although it still takes around 30 min to saturate. The blue shift, however, dominates first and takes a few seconds only to saturate, which can also be seen from Figure 2a. Therefore, it can easily be separated from the slower red shift, which results in both strong red and blue shifts, which is favorable for bidirectional trimming. Overall, a maximum refractive index increase of $2.4\times10^{-2}$ was observed with exposure to either the 405 nm or 520 nm laser, while the greatest decrease, $-2\times10^{-2}$, occurred with 520 nm laser exposure.
\\~\\
Figure 2d shows a different trimming behavior for $n_{srn}$ = 2.4. This time up to 3 different sections can be highlighted. The first section I, which corresponds to the lowest exposure powers, indicates only weak red shifts. With the increase of exposure power, in section II, we can first observe a weak blue shift which takes a few seconds to saturate, followed by a strong red shift, which in turn takes around 30 min to saturate. In case of exposure to 405 nm light, we also indicate section III where no blue shift could be observed, although a much stronger red shift can be seen. Remarkably, the maximum refractive index increase was 0.11, the largest ever reported, which indicates a great potential of this material for trimming. Meanwhile, the largest refractive index decrease was only $-0.5\times10^{-3}$ for both wavelengths of exposure.
\\~\\
We discuss the observed trend in greater detail in Supporting Information, section S4, where we conclude that thermal annealing explains the observed behavior. Specifically, a laser beam has a non-uniform distribution of power, which results in a temperature gradient within a section of an MRR. The sections of the ring heated to 350 °C – 550 °C induce the blue shifts, while the sections heated above 550 °C – the red shifts. Here we also briefly report that in case of index 2.9, 520 nm light is favored for bidirectional trimming since it causes lower temperature gradient in a SRN waveguide, which results in a stronger blue shift (see Supporting Information, section S4), while a stronger red shift can be achieved by simply increasing exposure power. Meanwhile, 405 nm is preferred when trimming $n_{srn}$ = 2.4 SRN MRRs since it causes a stronger red shift due to larger absorption and heating.
\\~\\
In summary, this complex behavior offers significant advantages for trimming. Both the direction and magnitude of the shifts, as well as the shift rates, can be finely tuned by adjusting the laser power, allowing higher power for coarse trimming and lower power for fine-tuning. This offers significant flexibility for trimming applications, depending on the specific requirements from a photonic device.

\subsection{Maximizing Resonance Shift}	

In the previous section we established that significant refractive index changes in both directions are possible. Therefore, in this section we aim to maximize the resonance shifts, and so we exposed the whole length of SRN MRRs to both 405 nm and 520 nm lasers. \textbf{Figure 3}a and Figure 3b exemplify the measured resonance shifts in SRN MRRs during exposure. Figure 3c demonstrates how resonances shift in the case of blue shift from Figure 3a. Meanwhile, Figure 3d and Figure 3e provide illustrations which demonstrate that the whole ring was exposed by steps. Specifically, the ring was split into even sections separated by the arc length $L_{step}$. Another parameter, $L_{exp}$, is the length of the exposed section of the ring, limited by a beam size. Given that the length of the ring was 170 $\mu$m, and the number of exposed sections was 18 – 30, based on Figure 3a and Figure 3b, $L_{step}$ was in a range of 5.5 $\mu$m – 9.5 $\mu$m. 
\\~\\
From Figure 3a, we observe a large blue shift of 10.6 nm and a red shift of 9.3 nm in index 2.9 SRN when exposed to 520 nm light. Meanwhile, from Figure 3b, a lower blue shift of 2.8 nm and the largest ever reported red shift of 49.1 nm were achieved in index 2.4 SRN when exposed to 405 nm light. Notably, blue shifts were saturated while the red shifts were not since each section of the MRR was exposed only for 10 min to expedite trimming process (it takes around 30 min for the red shift to saturate). Therefore, we conclude that the red shift can be even stronger. 
\\~\\
It is worth noting that depth of resonances changes after exposure, as apparent from Figure 3c. It is caused by change in transmission coefficient of a coupler section of a ring and increase in optical losses. The latter is due to scattering losses stemming from non-uniform refractive index profile caused by stepped exposures (Figure 3e). We demonstrated that a significant reduction in losses is possible if the ring is exposed more uniformly, and provided greater details in Supporting Information, section S5.
\\~\\
We can estimate the maximum resonance shift possible to achieve by using Equation (1-3) when the whole ring is exposed to saturation. First, we need to determine $L_{exp}$ from Equation 3, and for that we need to measure the resonance shift $\Delta \lambda_{sec}$ when only one section is exposed (Figure 1f), and $\Delta \lambda_{ring}$ when the whole ring is exposed (Figure 3d). Since the maximum blue shift of $\Delta \lambda_{ring}$ = 10.6 nm was measured in the case when $n_{srn}$ = 2.9, $\lambda_{exp}$ = 520 nm (Figure 3a), and $\Delta \lambda_{sec}$ = 540 pm blue shift was observed when only one section of the ring was exposed (Figure S2b), it was possible to estimate the exposure length: $L_{exp} \approx\;10 \mu m$. Now it is possible to estimate $\Delta \lambda_{ring}$ since $\Delta \lambda_{sec}$ is known. For example, when $n_{srn}$ = 2.4, $\lambda_{exp}$ = 405 nm, we determined that $\Delta \lambda_{sec}$ = 3900 pm (Figure S2c), and so $\Delta \lambda_{ring}$ = 66.3 nm. Both demonstrated $\Delta \lambda_{ring}$ = 49.1 nm and estimated $\Delta \lambda_{ring}$ = 66.3 nm are the largest among all known trimming methods to date, proving that SRN can be trimmed in a very high range.
\\~\\
\begin{figure}[htbp]
\includegraphics[width=1\linewidth]{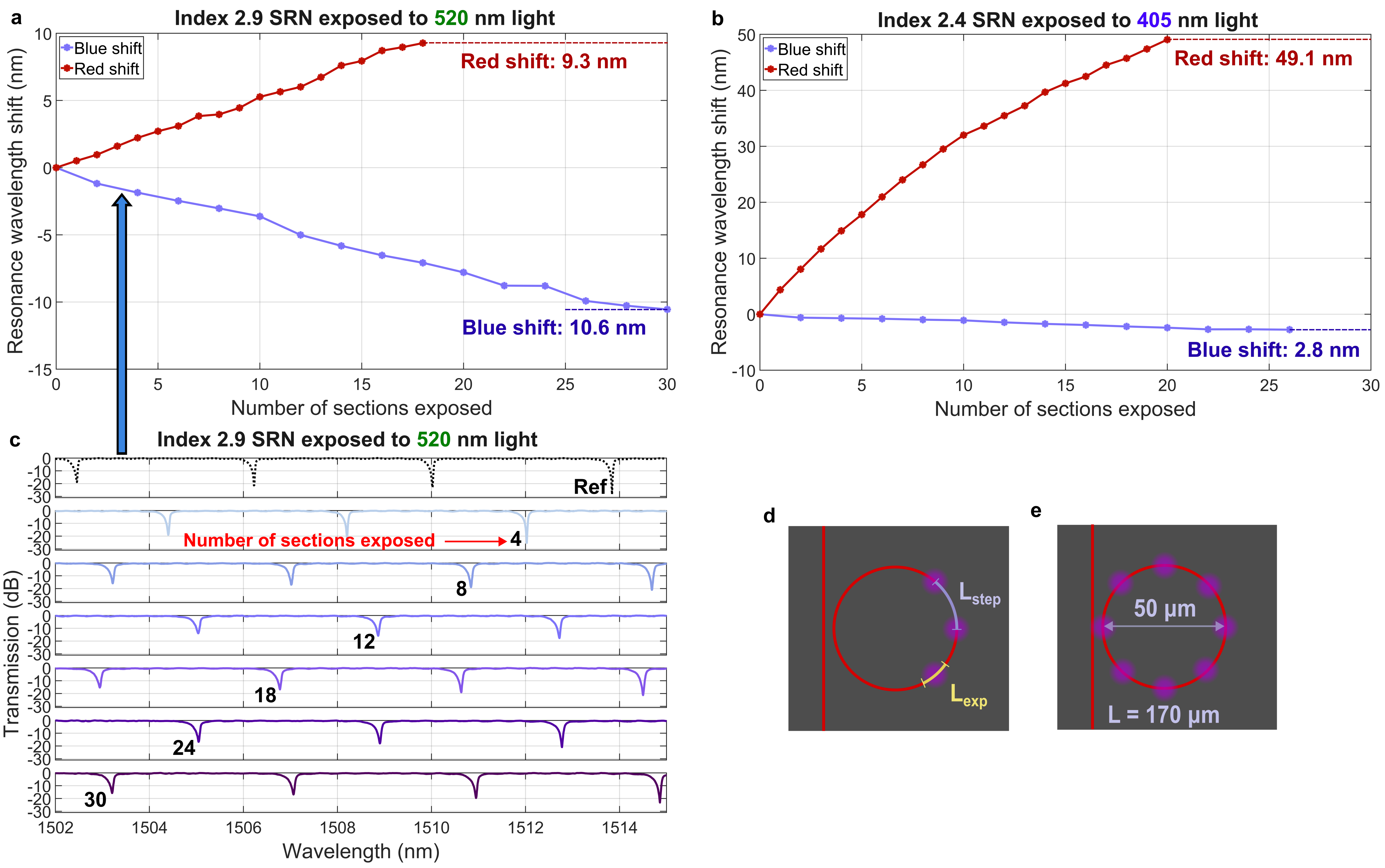}
\caption{Resonance wavelength shifts in SRN MRRs exposed to (a) 16.7 dBm of 520 nm light when $n_{srn}$ = 2.9 and (b) 16.0 dBm of 405 nm light when $n_{srn}$ = 2.4. (c) Transmission spectra of the probing signal in SRN MRR for the case of blue shift from Figure 3a. The dotted black curve indicates the initial position of a resonance (reference). The numbers next to the resonances indicate the number of sections exposed. (d) Illustration of how $L_{exp}$ and $L_{step}$ are defined. (e) Illustration of how the whole ring was exposed, where $L_{step}$ varied in a range of 5.5 $\mu$m – 9.5 $\mu$m based on number of sections exposed. 
\label{fig3}}
\end{figure}

\begin{table} [!ht]
\caption{Maximum estimated resonance shifts $\Delta \lambda_{ring}$ when the whole ring is exposed.}
\begin{tabular}{@{}llllllll@{}}
\hline
&$n_{srn}$ = 2.4, &$n_{srn}$ = 2.4,  &$n_{srn}$ = 2.9, &$n_{srn}$ = 2.9, \\
&$\lambda_{exp}$ = 405 nm &$\lambda_{exp}$ = 520 nm &$\lambda_{exp}$ = 405 nm &$\lambda_{exp}$ = 520 nm
\\ 
\hline
$\Delta \lambda_{ring}$ [nm], red shift		&66.3	&36.6		&10.9	&10.9		\\
$\Delta \lambda_{ring}$ [nm], blue shift	&3.1	&2.7		&6.1\textsuperscript{a)}	&10.6\textsuperscript{a)}		\\
$\Delta \lambda_{ring}$ [nm], shift span 	&69.4	&39.3		&17		&21.5		\\
\hline
\end{tabular}
\label{tab:Table1}
\parbox{\linewidth}{\footnotesize\textit{\textsuperscript{a)}Measured value.}}
\end{table}

\textbf{Table 1} summarizes the estimated maximum resonance shifts possible when the whole ring is exposed to saturation. We used the best results from Figure 2c and Figure 2d, where we measured refractive index changes for different exposure powers. The shift span is achieved by subtracting $\Delta \lambda_{ring}$ for blue shift from $\Delta \lambda_{ring}$ for red shift.
\\~\\
As can be seen in Table 1, overall, $n_{srn}$ = 2.4 SRN yields a much larger range of trimming than $n_{srn}$ = 2.9 SRN, although the blue shift is much weaker. In contrast, in case of $n_{srn}$ = 2.9 SRN, the blue shift can be as strong as 10.6 nm when SRN is trimmed by the 520 nm laser, which favors bidirectional trimming. Therefore, there is a tradeoff between these two cases. 
\\~\\
We also provide \textbf{Table 2}, which summarizes our results and compares them with those of other research groups. We focus on the best results achieved in our work and demonstrated elsewhere for any trimming method.
\\~\\

\begin{table}[!ht]
\caption{Summary of our experimental results compared to the results from other research groups, sorted by the wavelength shift span.}
\begin{tabular}{@{}llllllll@{}}
\hline
Ref. & $\Delta \lambda$ [nm], & $\Delta \lambda$ [nm], & $\Delta n_\text{srn} \times 10^{-2}$ & CMOS- & Material & Method \\
 & span & blue \& red & blue \& red & comp.? &  &  \\
  &  & shift & shift &  &  &  \\
\hline
25 & 2.5 & -2.5 & - & No & Chromophore-doped & Exposure to e-beam \\
 &  &  &  &  & polymer cladding &  \\
26 & 3 & -1 \& 2 & -0.4 \& 0.8 & No\textsuperscript{a)} & HSQ cladding & Laser annealing \\
24 & 4.9 & 4.9 & - & Yes & SiO\textsubscript{2} cladding & Exposure to e-beam \\
19 & 6.7 & 6.7 & 4.8 & No & As\textsubscript{2}S\textsubscript{3} cladding & Illumination with \\
 &  &  &  &  & & halogen lamp \\
41 & 7 & -7 & - & Yes & Si core implanted  & Laser annealing \\
 &  &  &  &  & with Ge ions &  \\
42 & 7 & -7 & -0.7 & Yes & Nitrogen-rich SiN core & Exposure to 244 nm light \\
39 & 8 & -8 & - & Yes & a-Si core & Exposure to 405 nm light \\
22 & 8.7 & -8.7 & - & Yes & c-Si core & Oxidizing by \\
 &  &  &  &  &  & 532 nm CW laser \\
17 & 9.7 & -9.7 & -7.5 & No & Polysilane cladding & Exposure to 370 nm light \\
18 & 12.1 & -12.1 & -1.3 & Yes & SiN core & Exposure to 244 nm light \\
23 & 16 & -10.5 \& 5.5 & - & Yes & c-Si core & Amorphization and ablation \\
 &  &  &  &  &  & by 400 nm femtosecond laser \\
\textbf{This work} & 19.9 & -10.6 \& 9.3 & -2 \& 2.4 & Yes & Index 2.9 SRN core & Annealing by 405 nm \\
&  &  &  &  &  & and 520 nm light \\
\textbf{This work} & 51.9 & -2.8 \& 49.1 & -0.5 \& 11.2 & Yes & Index 2.4 SRN core & Annealing by 405 nm \\
&  &  &  &  &  & and 520 nm light \\
\hline
\end{tabular}
\label{tab:Table2}
\parbox{\linewidth}{\footnotesize\textit{\textsuperscript{a)}When used as a cladding due to instability of optical properties caused by a change in humidity [26].}}
\end{table}

While some groups have successfully achieved strong blue shifts, inducing a substantial red shift has proven particularly challenging. The largest red shift reported so far, 6.7 nm, required a CMOS-incompatible chalcogenide cladding [19]. In contrast, our work demonstrates nearly an order of magnitude improvement in refractive index using SRN with a refractive index of 2.4. Furthermore, most studies report only unidirectional resonance shifts, with few exceptions. For instance, a bidirectional shift was achieved using HSQ cladding [26]; however, this approach suffered from significant backshifting due to moisture absorption [26]. Alternatively, bidirectional shifts in crystalline silicon (c-Si) have been demonstrated via ablation and amorphization using a 400 nm femtosecond laser (second harmonic of a Ti:Sa laser) [23]. This method, however, entails considerably higher trimming costs, limits precision, and was only demonstrated in uncladded samples [23], raising concerns about its compatibility with cladded devices. To clarify, trimming an uncladded device introduces uncertainty because the final resonance position after cladding is unpredictable. This undermines one of the primary advantages of trimming: achieving controlled and precise resonance adjustment. In summary, the bidirectional tuning capability, record-breaking red shift, CMOS-compatibility, and precise, cost-effective trimming achieved in SRN clearly underscore its exceptional performance.

\subsection{Resonance Backshifting in Index 2.9 SRN MRRs}

It is known that the resonance shift induced by trimming may drift over time [26]. Therefore, it was important to assess the stability of both red and blue shifts in our case. \textbf{Figure 4} displays the MRR resonance wavelength shift tracked over several days after exposure. In this work, we define backshifting as the resonance shift of an MRR moving in the direction opposite to the initial tuning shift. Conversely, forward shifting describes resonance movement continuing along the original shift direction.
\\~\\
In Figure 4a, a backshift (red shift in this case) of 240 pm was observed 35 days after exposure of the MRR where an initial blue shift of 5.5 nm was induced. This result aligns with previous reports, where long-term backshifting is commonly observed after trimming [26]. However, the backshift is considerably smaller than the initial 5.5 nm shift, making wide-range trimming approach suitable for applications where accuracy within approximately 200 pm is acceptable.
\\~\\
\begin{figure}[htbp]
\includegraphics[width=1\linewidth]{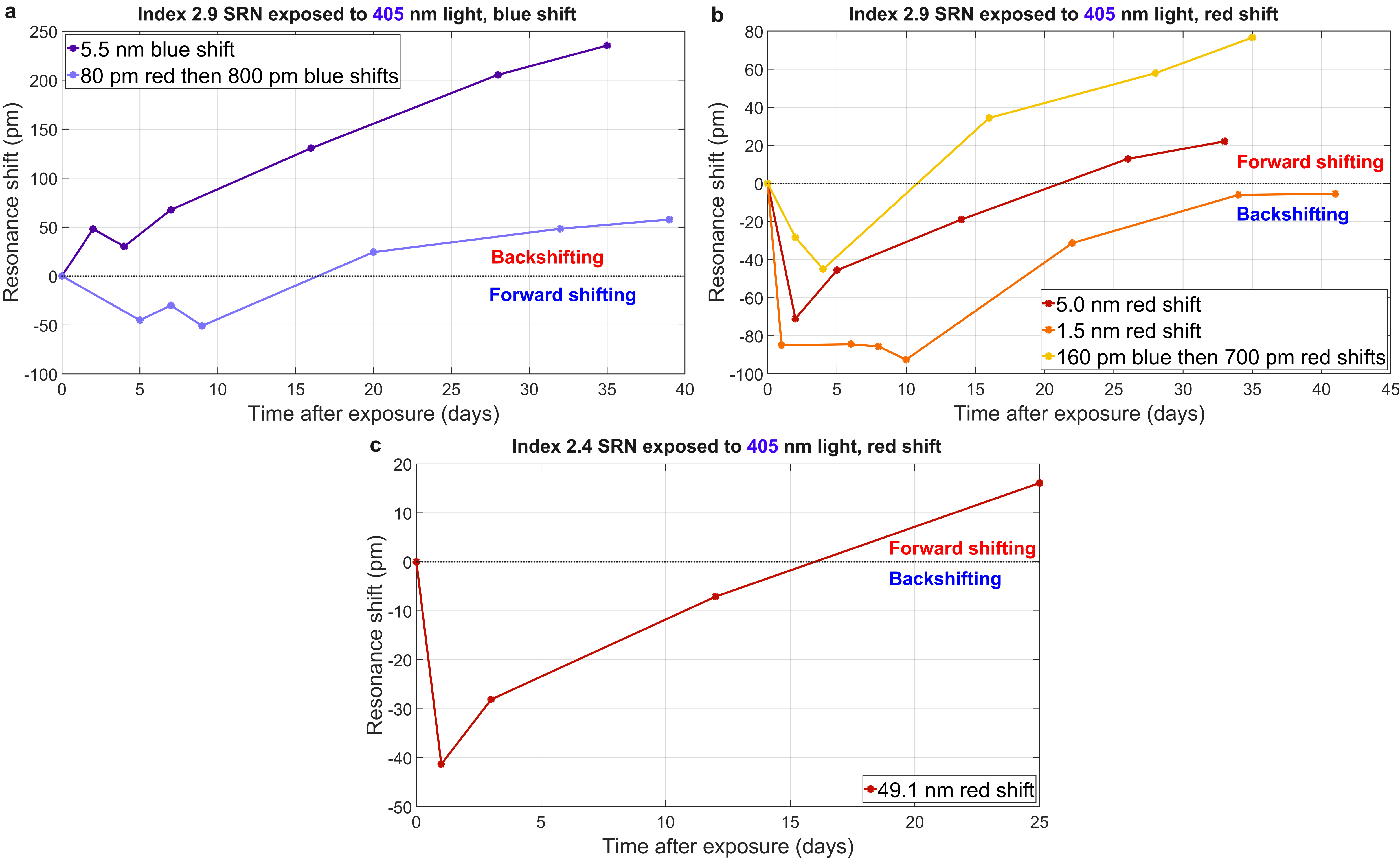}
\caption{Resonant wavelength shift as a function of time after exposure when MRRs are exposed to 405 nm light in case of (a) a blue shift, $n_{srn}$ = 2.9, (b) a red shift, $n_{srn}$ = 2.9, (c) a red shift, $n_{srn}$ = 2.4. The dotted black line separates the areas of forward shifting and backshifting. 
\label{fig4}}
\end{figure}

Additionally, in one case where an initial 80 pm red shift was followed by an 800 pm blue shift, a forward shift was observed within the first 10 days instead of the expected backshift. This suggests that the red shift dominated the early back-shifting process immediately after exposure. Over time, however, the blue shift prevailed, causing the resonance to shift in the opposite direction. This behavior suggests that red shifts tend to dominate shortly after exposure, while blue shifts have a more pronounced long-term influence. 
\\~\\
To further support this, in another example, shown in Figure 4b, where a 1.5 nm red shift was initially induced, the resonance first back-shifted within the first 10 days, after which it reversed the direction of shifting, and gradually returned close to its original position over 41 days. Similar dynamics was observed when the initial shift was 5.0 nm, where resonance returned close to its initial position after 33 days. This demonstrates that, under certain conditions, resonance backshifting can be almost fully mitigated when a balance between red and blue shifts is achieved. 
\\~\\
In addition, we also provide an example where a resonance was first blue-shifted by 160 pm, followed by a 700 pm red shift. In this case, the blue shift eventually prevailed, forward shifting the resonance by 80 pm without signs of saturation. This suggests that excessive blue shift may still cause significant (forward) shifts even after a red shift is induced post-exposure.
\\~\\
Finally, Figure 4c demonstrates that resonance relaxation dynamics at $n_{srn}$ = 2.4 is similar to that at $n_{srn}$ = 2.9. Notably, the observed instability was under 100 pm, despite an initial large red shift of 49.1 nm — highlighting the exceptional stability and suitability of SRN for trimming applications.
\\~\\
In conclusion, red and blue shifts coexist and drive resonance shifts in opposite directions, but they follow distinct dynamic patterns. Red shifts dominate within the first 3-10 days, while blue shifts tend to take over afterward. The stronger backshifting effects caused by blue shifts suggest that large blue shifts may be less desirable for long-term stability. However, a balanced combination of red and blue shifts can lead to resonance stabilization post-exposure. 

\section{Application Demonstration: Fine Trimming of a Demultiplexer} 

In this section, we demonstrate the benefits of bidirectional refractive index trimming. We adjust the passbands of a wavelength-division multiplexing (WDM) demultiplexer fabricated using SRN with a refractive index of $n_{srn}$ = 2.9 (at 1550 nm). This example illustrates that a passband can be trimmed with the accuracy of about 10 pm, proving the effectiveness of the approach.
\\~\\
\textbf{Figure 5}a presents a microscopic image of the WDM demultiplexer used in practical trimming experiments. In addition, Figure 5b shows a magnified image of an MRR that was trimmed. The ring was made of four straight sections connected by four arcs with the radius of 25 $\mu$m. The straight sections at the couplers’ sides had the length of 16 $\mu$m, while the length of the other two straight sections was 10 $\mu$m. In total, the length of the ring was about 209 $\mu$m.
\\~\\
Figure 5c and Figure 5d show the transfer characteristics of the WDM demultiplexer at a through port and one of drop ports, respectively, before and after trimming with a 405 nm laser. The objective of the trimming process was to set the passbands of the demultiplexer at 1550 nm, 1549.25 nm, and 1548.5 nm such that they are separated by 750 pm. Based on Figure 5c and Figure 5d, we achieved the trimming accuracy of about 10 pm. Indeed, the final resonances’ positions were 1548.508 nm, 1549.248 nm, and 1549.988 nm, which means they were away from the targets by 8 pm, 2 pm, and 12 pm, respectively. Notably, the trimming technique enabled both red and blue shifts of resonances, providing a versatile method for fine-tuning the device’s filtering properties. The switching of the direction and the rate of resonance shifting was enabled by changing the laser’s power.
\\~\\

\begin{figure}[htbp]
\includegraphics[width=1\linewidth]{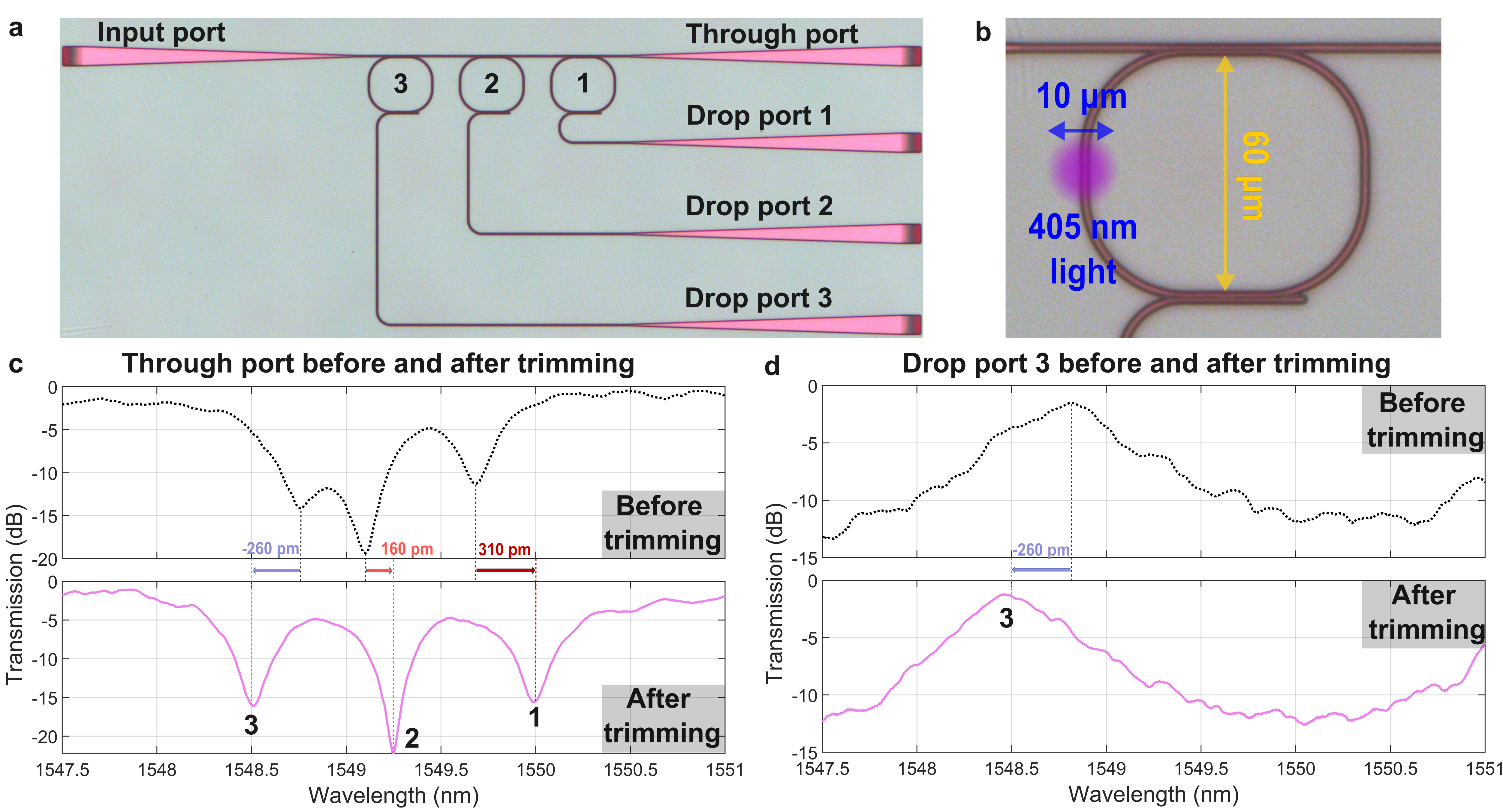}
\caption{(a) Microscope image of a fabricated demultiplexer. (b) Magnified microscope image of an MRR. Transmission of the probing signal in SRN MRRs as a function of the wavelength for (c) through port and (d) drop port 3. The values above the arrows represent resonance shifts necessary for trimming to achieve the target. The numbers 1,2,3 represent each ring from Figure 5a.
\label{fig5}}
\end{figure}

Here we highlight the importance of bidirectional shifting. Indeed, the initial position of the resonance "3" (from Figure 5c) needed to be blue-shifted by 260 pm, while the position of the resonance "2" – red-shifted by only 160 pm, and finally, a stronger red shift of 310 pm was needed for the resonance "1". If only blue shift was available, the resonances "1" and "2" would need to be blue-shifted by FSR - $\Delta \lambda_{red}$, where $\Delta \lambda_{red}$ is the resonance red shift needed. Given that typical FSR is in a range between 1 nm and 5 nm, that would require a much stronger blue shift from 850 pm to 4850 pm for the resonance "1". A large shift is undesirable since it will lead to stronger backshifting and increased losses. 
\\~\\
Thus, the results confirm that the trimming approach is suitable for sophisticated photonic devices, providing the necessary precision for fine-tuning passbands in WDM applications. By leveraging the bidirectional trimming method, we can achieve enhanced performance in SRN-based integrated photonic systems.

\section{Discussion} 

Our study demonstrates the remarkable potential of SRN as a reconfigurable photonics platform, offering unprecedented control over the refractive index through visible light trimming. The ability to achieve bidirectional resonance shifts, both blue and red, with a single laser source represents a significant advancement over most existing trimming methods, which are largely limited to unidirectional index changes. This capability opens new possibilities for dynamic tuning of phase-sensitive devices.
\\~\\
The observed resonance shifts, reaching up to 10.6 nm for blue shifts and a record-setting 49.1 nm for red shifts, underscore SRN’s scalability and versatility across diverse photonic applications. These shifts correspond to refractive index changes of approximately $-2\times10^{-2}$ and 0.11, respectively, which are not only substantial but also finely tunable through adjustments in laser power and exposure duration. This tunability addresses a critical limitation in photonic device manufacturing by providing a means to precisely control device performance post-fabrication, effectively compensating for inherent variations.
\\~\\
Moreover, the coexistence of red and blue shifts provides an additional benefit: significantly reduced trimming backshifting over time, thus contributing to the long-term stability of the trimmed devices. This stabilization is particularly valuable for photonic devices where phase accuracy is critical, such as in WDM systems and other high-performance optical networks.
\\~\\
Our results also highlight the cost-effectiveness of this trimming approach. Unlike alternative methods that often require CMOS-incompatible materials, or expensive trimming methods, our approach uses widely available equipment and CMOS-compatible materials, positioning it as a practical solution for scaled up and industrial applications.
\\~\\
Additionally, the bidirectional shifting in SRN arises from its thermal annealing properties. We established that at lower annealing temperatures (400 °C – 500 °C), the refractive index decreases, while above 500 °C – 600 °C, it increases. This behavior aligns with trimming trends, where lower power induces blue shifts and higher power causes red shifts. Also, refractive index reduction is more pronounced in higher-index SRN during rapid thermal annealing, matching observations in trimming, where higher-index SRN undergoes larger blue shifts. Furthermore, thermal simulations revealed that SRN can be heated to temperatures exceeding 1000 °C using only 16 dBm of CW 405 nm laser radiation. This is attributed to its exceptionally low thermal conductivity, which facilitates effective heat localization during laser trimming. Therefore, the simulation results confirm that it is possible to induce both red and blue shifts during trimming. Overall, these findings suggest that trimming behavior can be effectively predicted by thermal simulation and annealing SRN thin films at different temperatures and measuring refractive index changes, which is a much faster and cheaper alternative to fabricating SRN waveguides.
\\~\\
As it has been discussed previously, subjecting SRN to rapid thermal annealing (RTA) at elevated temperatures (typically above 600 °C) results in the dissociation of Si-H bonds, substantial hydrogen (H) desorption, a reduction in nitrogen (N) content, and an increase in the concentration of Si-Si [34-36]. On the other hand, studies indicate that RTA at lower temperatures (below 600 °C) promotes bond redistribution (Si-Si, Si-N, Si-H, N-H), leading to a rise in Si-H and Si-N bond concentrations while decreasing N-H and Si-Si bond concentrations [34,35]. In our study, we also observed a noticeable change in behavior when RTA temperatures exceeded 600 °C. Based on these findings, we hypothesize that the refractive index reduction at \textless 600 °C is due to bond redistribution which results in increased Si-H and Si-N bond formation, while the refractive index increase at \textgreater 600 °C results from significant H desorption due to Si-H bond dissociation, and nitrogen reduction.
\\~\\
While both 405 nm and 520 nm lasers enable bidirectional trimming as both wavelengths are absorbed by SRN, generating heat, it is important to acknowledge that 520 nm light is preferred for index 2.9 SRN bidirectional trimming since it causes a stronger blue shift. It was demonstrated that this is due to lower absorption resulting in lower temperature gradient, which was proven to favor blue shift. Meanwhile, in SRN with a refractive index of 2.4, red shifts are noticeably smaller under 520 nm exposure due to its much lower absorption (extinction coefficients are 0.44 for 405 nm and 0.08 for 520 nm), resulting in less heating, so 405 nm light trimming is preferred.
\\~\\
We also would like to compare our trimming approach to widely used heaters. First, the primary advantage of fiber-based laser trimming is cost efficiency. The one-time cost of setting up the trimming process—including fiber and diode lasers—is minimal compared to the recurring fabrication costs of heaters. Furthermore, heaters require continuous energy input to maintain resonance tuning, whereas laser trimming is a one-time adjustment (e.g., to set the resonance wavelength of a microring resonator), after which no further energy is needed. Another key benefit of laser trimming is its bidirectional tuning capability, unlike heaters, which are generally limited to red-shifting. Additionally, packaging of trimming fibers is not required since trimming is performed only once (e.g., by scanning the trimming fiber's position), and the fibers can be removed afterward.
\\~\\
On the other hand, heaters excel in scenarios requiring large-scale integration due to their scalability. They allow for real-time tuning of resonance positions, enabling dynamic modulation or switching. Moreover, heater technology is well-established and widely adopted in CMOS foundries, making it a reliable and standardized solution for integrated photonics.
\\~\\
Ultimately, both approaches have their practical uses and complement each other rather than compete. In fact, combining the two methods can provide an optimal solution. For instance, laser trimming can be employed to permanently adjust the resonance position close to its target value, while heaters can be used for fine-tuning, modulation, or switching. This combination significantly reduces the power required for heaters, as they only need to compensate for small perturbations caused by environmental factors, such as temperature fluctuations. Thus, integrating both methods enhances the overall functionality and efficiency of photonic devices.
\\~\\
Finally, we identify new applications for SRN in integrated photonics. For example, SRN shows promise as a CMOS-compatible coating for silicon-on-insulator (SOI) platforms, enabling bidirectional tuning through visible light exposure. Its capacity for a significant refractive index increase also suggests potential for creating etch-free SRN waveguides using visible light, which could reduce fabrication costs and minimize scattering losses due to the absence of etched sidewalls.
\\~\\
In conclusion, this work establishes visible light trimming of SRN as a powerful, flexible tool for photonic device optimization. The demonstrated bidirectional tuning capability, coupled with its precision, cost-effectiveness, and stability, positions SRN as a key material for the next generation of integrated photonics. Future research should explore extending this technique to other photonic architectures, potentially broadening the range of applications and further refining the precision and scalability of SRN-based devices.

\section{Materials and Fabrication Process} 

The fabrication process of PECVD SRN MRRs began with the removal of the silicon device layer from silicon-on-insulator (SOI) substrates using tetramethylammonium hydroxide (TMAH) wet etching, leaving only a 3 $\mu$m buried oxide (BOX) layer on top of a 700 $\mu$m Si substrate (note that this step can be skipped if Si samples with only SiO\textsubscript{2} layer on top are available). Next, PECVD SRN films were deposited in an Oxford Plasmalab PECVD system at a temperature of 350 °C and a pressure of 650 mTorr. The SiH\textsubscript{4} and N\textsubscript{2} precursor gases were introduced at flow rates of 80 sccm and 800 sccm, respectively, for the 2.4 refractive index film, and 450 sccm and 200 sccm for the 2.9 refractive index film. Following deposition, hydrogen silsesquioxane (HSQ) e-beam resist was spun onto the SRN films, and electron-beam lithography was used to define the waveguide patterns. After developing the HSQ resist, the SRN films were etched in a Plasma-Therm ICP-RIE system, and the HSQ was subsequently removed using a 6:1 buffered oxide etch (BOE) diluted with deionized water in ratio 1:20. Complete removal of the HSQ resist was confirmed by measuring the remaining waveguide thickness with a stylus profilometer. The samples were then clad with a 2 $\mu$m PECVD SiO\textsubscript{2} layer, also deposited using the Oxford Plasmalab PECVD system.

\medskip
\textbf{Acknowledgements} \par 
We thank SDNI and all UCSD's nano3 cleanroom staff and Dr Maribel Montero for their assistance with samples fabrication.

\medskip
\textbf{Funding} \par
This work was supported by the National Science Foundation (NSF) grants ECCS-2217453 and NSF ECCS-2410053, the San Diego Nanotechnology Infrastructure (SDNI) supported by the NSF National Nanotechnology Coordinated Infrastructure (grant ECCS-2025752), and the ASML/Cymer Corporation.

\medskip
\textbf{Author contributions} \par
D.B. assembled a setup, performed most experiments, processed data, determined the direction of research. M.D. fabricated samples and assisted with experiments. K.J. assisted with experiments, manuscript preparation, exchanged ideas. V.F. and A.G. assisted with manuscript preparation, discussed ideas. N.A., A.N., P.K.L.Y. and Y.F. assisted with manuscript preparation, exchanged ideas, provided funds and equipment.

\medskip
\textbf{Conflict of interest} \par
The authors declare no conflicts of interest regarding this article.

\medskip
\textbf{Data availability statement} \par
The datasets generated during and/or analyzed during the current study are available from the corresponding author on reasonable request.

\medskip
\textbf{References} \par

[1] D.~Belogolovskii, N.~Alic, A.~Grieco, Y.~Fainman, 
\newblock \emph{Adv. Photonics Res.} \textbf{2024}, \emph{5}, 2400017.

[2] D.~T.~H.~Tan, D.~K.~T.~Ng, J.~W.~Choi, E.~Sahin, B.-U.~Sohn, 
\newblock \emph{Adv. Phys.: X} \textbf{2021}, \emph{6}, 1905544.

[3] D.~Belogolovskii, Y.~Fainman, N.~Alic, 
\newblock \emph{Adv. Opt. Mater.} \textbf{2024}, \emph{12}, 2401299.

[4] H.~Nejadriahi, S.~Pappert, Y.~Fainman, 
\newblock \emph{Opt. Express} \textbf{2020}, \emph{28}, 24951.

[5] K.~Ooi, D.~Ng, T.~Wang, A.~Chee, S.~Ng, 
\newblock \emph{Nat. Commun.} \textbf{2017}, \emph{8}, 13878.

[6] A.~Friedman, H.~Nejadriahi, R.~Sharma, Y.~Fainman, 
\newblock \emph{Opt. Lett.} \textbf{2021}, \emph{46}, 17, 4236.

[7] G.-R.~Lin, S.-P.~Su, C.-L.~Wu, Y.-H.~Lin, B.-J.~Huang, 
\newblock \emph{Sci. Rep.} \textbf{2015}, \emph{5}, 9611.

[8] H.~H.~Lin, K.~S.~Wong, A.~F.~Kim, R.~Takahashi, 
\newblock \emph{APL Photonics} \textbf{2019}, \emph{4}, 036101.

[9] C.~Lacava, S.~Pappert, I.~Cristiani, P.~Minzioni, 
\newblock \emph{Photon. Res.} \textbf{2019}, \emph{7}, 615.

[10] J.~Choi, J.~Chen, G.~F.~R.~Ng, 
\newblock \emph{Sci. Rep.} \textbf{2016}, \emph{6}, 27120.

[11] A.~Friedman, D.~Belogolovskii, A.~Grieco, Y.~Fainman, 
\newblock \emph{Opt. Express} \textbf{2022}, \emph{30}, 45340.

[12] H.~Nejadriahi, S.~Pappert, Y.~Fainman, P.~Yu, 
\newblock \emph{Opt. Lett.} \textbf{2021}, \emph{46}, 4646.

[13] Z.~Lu, J.~Shen, M.~He, J.~Yao, 
\newblock \emph{Opt. Express} \textbf{2017}, \emph{25}, 9712.

[14] D.~S.~Boning, S.~I.~El-Henawy, Z.~Zhang, 
\newblock \emph{J. Lightwave Technol.} \textbf{2022}, \emph{40}, 1776.

[15] W.~A.~Zortman, D.~C.~Trotter, M.~R.~Watts, 
\newblock \emph{Opt. Express} \textbf{2010}, \emph{18}, 23598.

[16] K.~O.~Hill, Y.~Fujii, D.~C.~Johnson, B.~S.~Kawasaki, 
\newblock \emph{Appl. Phys. Lett.} \textbf{1978}, \emph{32}, 647.

[17] S.~Suzuki, Y.~Hatakeyama, Y.~Kokubun, S.~T.~Chu, 
\newblock \emph{J. Lightwave Technol.} \textbf{2002}, \emph{20}, 745.

[18] H.~Haeiwa, T.~Naganawa, Y.~Kokubun, 
\newblock \emph{IEEE Photonics Technol. Lett.} \textbf{2004}, \emph{16}, 135.

[19] A.~Canciamilla, S.~Vallaitis, T.~Dangel, D.~Bogaerts, 
\newblock \emph{Opt. Express} \textbf{2012}, \emph{20}, 15807.

[20] T.~Guo, M.~Zhang, Y.~Yin, D.~Dai, 
\newblock \emph{IEEE Photonics Technol. Lett.} \textbf{2017}, \emph{29}, 419.

[21] J.~Zheng, C.~Wu, M.~Shen, Z.~Li, 
\newblock \emph{Opt. Mater. Express} \textbf{2018}, \emph{8}, 1551.

[22] C.~J.~Chen, J.~T.~Lin, Y.~C.~Wang, L.~Chang, 
\newblock \emph{Opt. Express} \textbf{2011}, \emph{19}, 12480.

[23] D.~Bachman, J.~Smith, Y.~Liang, 
\newblock \emph{Opt. Lett.} \textbf{2011}, \emph{36}, 4695.

[24] J.~Schrauwen, D.~Van Thourhout, R.~Baets, 
\newblock \emph{Opt. Express} \textbf{2008}, \emph{16}, 3738.

[25] S.~Prorok, A.~Y.~Petrov, M.~Eich, J.~Luo, 
\newblock \emph{Opt. Lett.} \textbf{2012}, \emph{37}, 3114.

[26] S.~Spector, J.~M.~Knecht, P.~W.~Juodawlkis, 
\newblock \emph{Opt. Express} \textbf{2016}, \emph{24}, 5996.

[27] D.~E.~Hagan, B.~Torres-Kulik, A.~P.~Knights, 
\newblock \emph{IEEE Photonics Technol. Lett.} \textbf{2019}, \emph{31}, 1373.

[28] X.~Yu, J.~Smith, R.~Zhang, 
\newblock \emph{IEEE Int. Conf. Group IV Photonics} \textbf{2019}, 1.

[29] X.~Yu, Z.~Li, Y.~Huang, 
\newblock \emph{Proc. SPIE} \textbf{2020}, \emph{11285}, 1128512.

[30] H.~Jayatilleka, Z.~Lu, W.~He, J.~Lin, 
\newblock \emph{J. Lightwave Technol.} \textbf{2021}, \emph{39}, 5083.

[31] Y.~Xie, H.~C.~Frankis, J.~D.~B.~Bradley, A.~P.~Knights, 
\newblock \emph{Opt. Mater. Express} \textbf{2021}, \emph{11}, 2401.

[32] S.~Lambert, W.~De Cort, J.~Beeckman, K.~Neyts, 
\newblock \emph{Opt. Lett.} \textbf{2012}, \emph{37}, 1475.

[33] S.~K.~Selvaraja, J.~Snyder, E.~Rosseel, 
\newblock \emph{Proc. IEEE Int. Conf. Group IV Photonics} \textbf{2011}, 71.

[34] S.~S.~He, V.~L.~Shannon, 
\newblock \emph{Proc. Int. Conf. Solid-State IC Technol.} \textbf{1995}, 269.

[35] F.~L.~Martínez, A.~del Prado, I.~Martil, 
\newblock \emph{Phys. Rev. B} \textbf{2001}, \emph{63}, 245320.

[36] P.~Dani, M.~Tuchen, B.~E.~Meli, J.~Franz, 
\newblock \emph{Micro Nano Eng.} \textbf{2024}, \emph{100}, 100291.

[37] M.~Hadi, S.~Pailhès, R.~Debord, A.~Benamrouche, E.~Drouard, T.~Gehin, C.~Botella, J.-L.~Leclercq, P.~Noe, F.~Fillot, V.~M.~Giordano,
\newblock \emph{Materialia} \textbf{2022}, \emph{26}, 101574.

[38]M.~Jacques, A.~Samani, E.~El-Fiky, D.~Patel, Z.~Xing, D.~V.~Plant,
\newblock \emph{Opt. Express} \textbf{2019}, \emph{27}, 10456.

[39] T.~Lipka, M.~Kiepsch, H.~K.~Trieu, J.~Müller, 
\newblock \emph{Opt. Express} \textbf{2014}, \emph{22}, 12122.

[40] J.~J.~Ackert, Z.~Liu, D.~X.~Yu, 
\newblock \emph{Opt. Express} \textbf{2011}, \emph{19}, 11969.

[41] M.~M.~Milosevic, X.~Chen, W.~Cao, A.~F.~Runge, 
\newblock \emph{IEEE J. Sel. Top. Quantum Electron.} \textbf{2018}, \emph{24}, 1.

[42] G.~De Paoli, J.~Smith, F.~Teller, 
\newblock \emph{Photon. Res.} \textbf{2020}, \emph{8}, 677.

\clearpage
\section*{Supplementary Information}
\renewcommand{\thesection}{S\arabic{section}} 
\setcounter{section}{0} 

\renewcommand{\thefigure}{S\arabic{figure}} 
\setcounter{figure}{0} 

\section{Aligning procedure of exposure fiber}

\begin{figure}[htbp]
\includegraphics[width=1\linewidth]{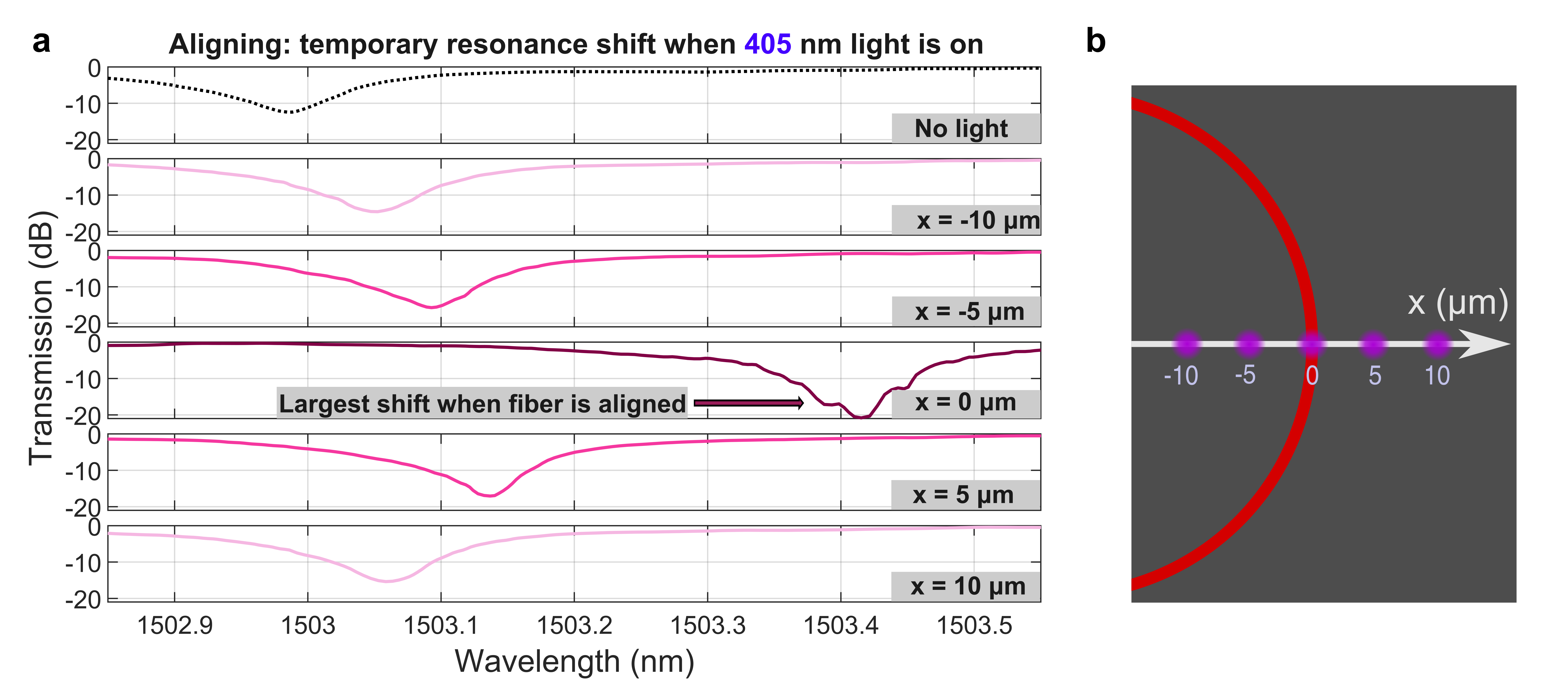}
\caption{(a) MRR transmission spectra when 405 nm exposing light (8.8 dBm) is off and on, plotted for multiple fiber offsets x. The dotted black curve represents the case when the exposing laser is off. The magenta/pink curves represent the cases when the 405 nm laser is on, and darker tint corresponds to a stronger wavelength shift. When x = 0, the fiber is perfectly centered over a section of the MRR. (b) Section of an MRR showing x-axis and the fiber offsets x from -10 $\mu$m to 10 $\mu$m used in (e).
\label{figS1}}
\end{figure}

Figure S1a and S1b clarify the aligning process. Specifically, Figure S1a illustrates how MRR transmission spectrum shifts as the fiber is moved from x = -10 $\mu$m to x = 10 $\mu$m. Meanwhile, Figure S1b demonstrates how x-axis is defined. The alignment of the exposing fiber was performed in two steps: coarse alignment using a camera microscope and fine alignment based on the temporary resonance shift caused by light absorption in SRN. The resonance red shift occurred as the fiber approached the MRR, with the maximum shift of 400 pm reached when the fiber was centered over the ring (x = 0 in Figure S1a and S1b). During alignment, low-power laser exposure was used to minimize permanent trimming, as low powers induce mostly temporary shifts due to heating and free carrier generation. It is worth noting that in practice, before alignment, the probing light was set at the resonance wavelength of the MRR, allowing resonance shifts to be tracked in real-time through changes in optical power. This approach enabled rapid fine alignment even with manual stages.

\section{Resonance shifting due to trimming of a section of an MRR}

Figure S2 exemplifies the measured transmission spectra of the probing signal in SRN MRRs when exposed to 405 nm and 520 nm light. The exposing fiber was centered over a section of the MRR (x = 0), as illustrated in Figure S1a and Figure S1b. Only one section of the MRR was exposed.
\\~\\

\begin{figure}[ht]
\includegraphics[width=1\linewidth]{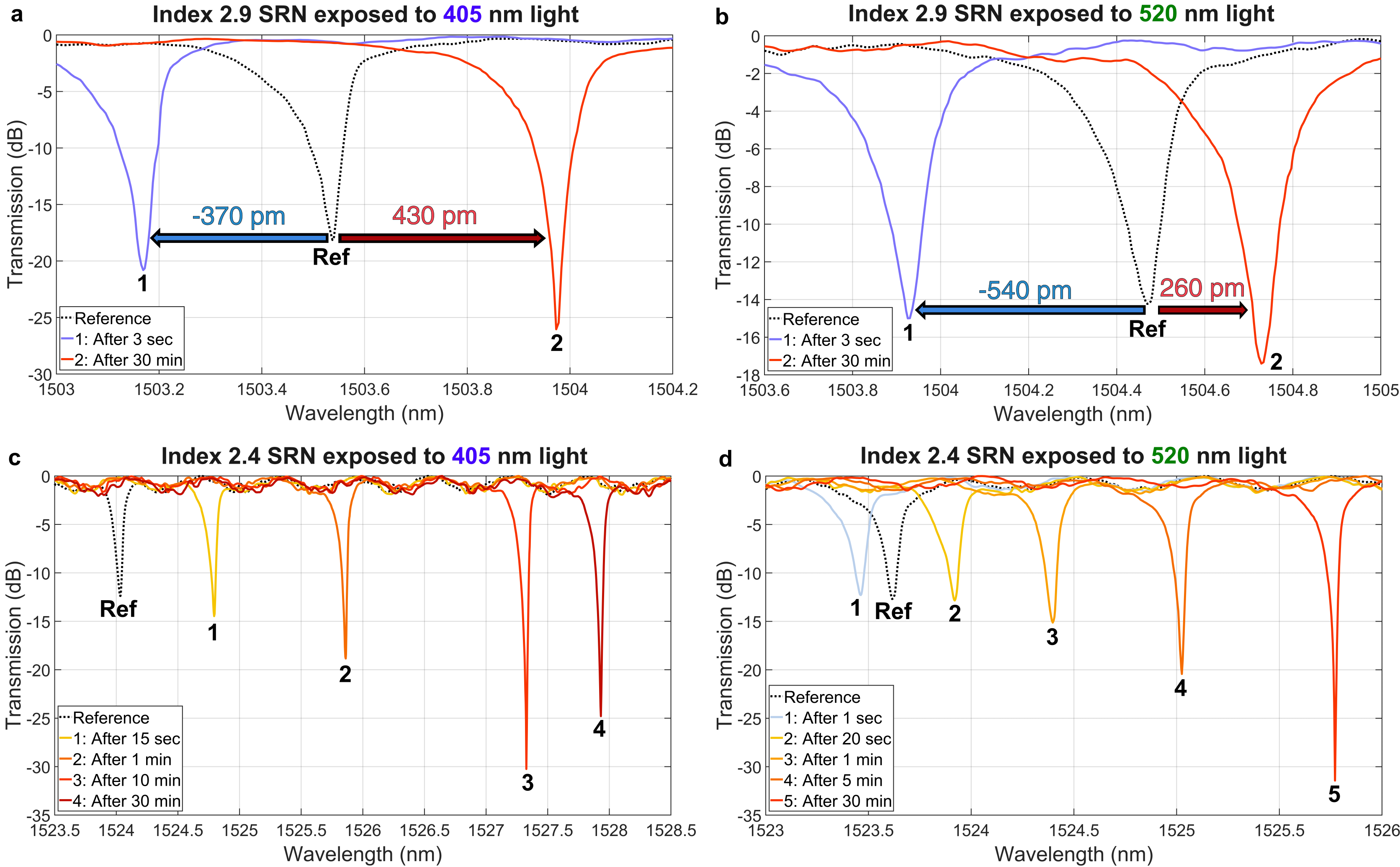}
\caption{Transmission spectra of the probing signal in SRN MRRs after trimming when (a) $n_{\text{srn}} = 2.9$, $\lambda_{\text{exp}} = 405$ nm, (b) $n_{\text{srn}} = 2.9$, $\lambda_{\text{exp}} = 520$ nm, (c) $n_{\text{srn}} = 2.4$, $\lambda_{\text{exp}} = 405$ nm, (d) $n_{\text{srn}} = 2.4$, $\lambda_{\text{exp}} = 520$ nm. The dotted black curve indicates the initial position of resonance (reference). The numbers (and “Ref”) next to the resonances indicate the order in which the spectra were measured (starting with “Ref”). The labels describe exposure time.
\label{figS2}}
\end{figure}

The results in Figure S2 confirm that it is indeed possible to achieve both red and blue resonance shifts using a single laser source ($P_{\text{exp}} = 16.0$ dBm). A blue shift was observed within a few seconds of exposure in every case except when $n_{\text{srn}} = 2.4$, $\lambda_{\text{exp}} = 405$ nm (blue shift is still possible at lower power), followed by a reversal of shift direction. In contrast, the red shift took up to 30 minutes to saturate. These results demonstrate the ability to clearly separate blue and red shifts by varying the duration of exposure, a crucial property for trimming applications. 
\\~\\
Noticeably, for $n_{\text{srn}} = 2.4$, the red shift was significantly stronger than for $n_{\text{srn}} = 2.9$. Specifically, the red shift was 430 pm for $n_{\text{srn}} = 2.9$, $\lambda_{\text{exp}} = 405$ nm, but as much as 3900 pm for $n_{\text{srn}} = 2.4$, $\lambda_{\text{exp}} = 405$ nm. Meanwhile, the strongest blue shift was observed when $n_{\text{srn}} = 2.9$, $\lambda_{\text{exp}} = 520$ nm.

\section{Thermal annealing as the origin of the bidirectional shifting in SRN MRRs}

In this section we demonstrate that thermal annealing is a cause of the bidirectional resonance shifting observed in SRN MRRs. We investigate how temperature of rapid thermal annealing (RTA) affects the refractive index (RI) change of SRN thin film. The refractive index of SRN ($n_{\text{srn}}$) films and their thicknesses are measured by ellipsometer. Figure S3a presents nsrn change as a function of RTA temperature after 2 min of RTA, while Figure S3b also demonstrates results for 6 min annealing. 

\begin{figure}[htbp]
\includegraphics[width=1\linewidth]{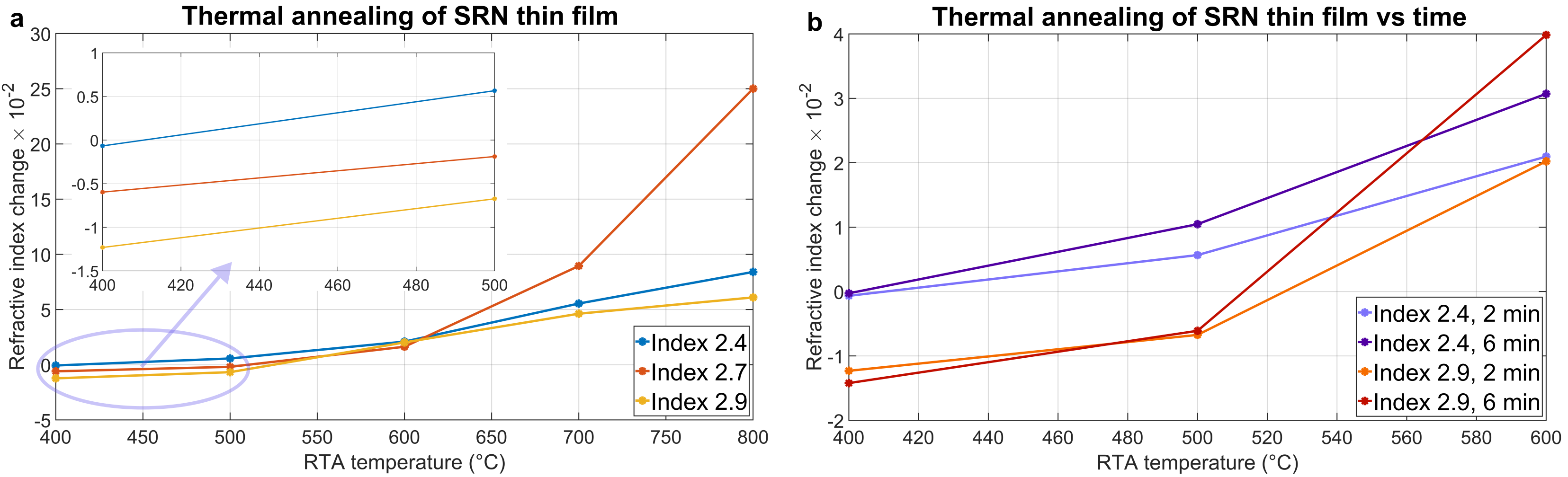}
\caption{(a) SRN refractive index change as a function of RTA temperature. The inset represents an area of a plot with RTA temperatures in range 400 °C – 500 °C. (b) SRN refractive index change as a function of RTA temperature for different times of RTA.
\label{figS3}}
\end{figure}

Based on Figure S3, it is evident that RTA is a cause of the bidirectional shifting observed in SRN. First, we can see a decline in $n_{\text{srn}}$ when annealing is done at lower temperatures. Specifically, the refractive index (RI) reduction is observed at 400 °C RTA for all $n_{\text{srn}}$ values of 2.4, 2.7, 2.9, and for $n_{\text{srn}}$ values of 2.7 and 2.9, the RI reduces even at 500 °C RTA. Importantly, once the RTA temperature exceeds 600 °C, $n_{\text{srn}}$ increases in all cases. This matches the behavior of the MRRs' resonance shifting, where lower exposure power induces a blue shift (due to less heating), while larger power causes a red shift. For example, the RI decrease in $n_{\text{srn}} = 2.9$ due to RTA is $1.4 \times 10^{-2}$, which is of the same order compared to the RI decrease measured during trimming experiments ($2 \times 10^{-2}$). The maximum RI increase of $2.4 \times 10^{-2}$ during trimming corresponds to RTA at 600 °C and above. 
\\~\\
Second, the RI reduction is larger in SRN with a higher $n_{\text{srn}}$, as can be seen from the inset provided in Figure S3a. This matches perfectly the trend observed during trimming of $n_{\text{srn}} = 2.4$ and $n_{\text{srn}} = 2.9$ SRN, where the blue shift was considerably stronger in $n_{\text{srn}} = 2.9$. The RI increase, on the other hand, appears larger in $n_{\text{srn}} = 2.4$, which also agrees well with the trimming results.
\\~\\
In addition, the trend observed from trimming (Figure 2c, Figure 2d) indicates that 520 nm light requires higher power to achieve a strong red shift. This is to be expected since 520 nm light is absorbed less by SRN, which results in less heating. For example, the extinction coefficients are 0.44 for 405 nm (penetration depth – 73 nm) and 0.08 (penetration depth – 530 nm) for 520 nm light in $n_{\text{srn}} = 2.4$ SRN. Therefore, it is expected that 405 nm light should cause a stronger red shift. Given that the thickness of index 2.4 SRN waveguides is about 380 nm, we conclude, based on the Beer–Lambert law, that only 50\% of all 520 nm light is absorbed (although the Si substrate absorbs the rest, causing some extra heating). This explains why 520 nm light requires about 2 dB more power to induce the same RI change (Figure 2d).

\section{On the origin of the bidirectional shifting: exposure power dependence}

In this section, we provide a more detailed explanation of the trend observed in Figure 2c. Since the trimming is performed with a laser beam that has a non-uniform power distribution, a temperature gradient is created within a section of an MRR. As a result, different regions of the MRR simultaneously experience both increases and decreases in refractive index during exposure. This effect is particularly pronounced in the index 2.9 SRN, where strong red and blue shifts were observed. Therefore, we focus on this case here.
\\~\\
Figure S4a shows how the dynamics change at different exposure powers when $n_{\text{srn}} = 2.9$ and $\lambda_{\text{exp}} = 520$ nm. Meanwhile, Figure S4b (identical to Figure 2c, shown here for convenience) depicts the refractive index (RI) change as a function of exposure power for $n_{\text{srn}} = 2.9$. To understand the behavior in Figure S4b, we include Figure S4c (arc exposure) and Figure S4d, where we discuss how temperature gradients, caused by non-uniform power distribution, affect trimming.
\\~\\
We start with case I, where power is low enough that even the peak temperature remains below 350°C (the SRN deposition temperature), resulting in negligible shifts. As power increases in section II, we observe a slower blue shift since parts of the exposed MRR section are heated between 350°C and 600°C. This is illustrated in Figure S4a, where only a blue shift is observed at $P_{\text{exp}} = 13$ dBm, saturating within 7 minutes (hence we call it ‘slow’). In section III, with further power increase, some ring sections reach temperatures above 600°C, where a red shift dominates, while others remain between 350°C and 600°C, where a blue shift persists. This results in bidirectional shifting. However, because both shifts are slow (5 to 30 minutes), they may partially offset each other, reducing the overall trimming range. Finally, in section IV, at the highest power levels, a stronger red shift is observed, though a blue shift is still present. At this stage, most of the ring section reaches temperatures above 600°C, which intensifies the red shift. Notably, the blue shift dynamics accelerate significantly, with saturation occurring in about 10 seconds (as seen in Figure S4a for $P_{\text{exp}} = 15.7$ dBm). In contrast, the red shift saturates in about 30 minutes, allowing red and blue shifts to be separated by exposure time, which is essential for bidirectional trimming.
\\~\\
\begin{figure}[htbp]
\includegraphics[width=1\linewidth]{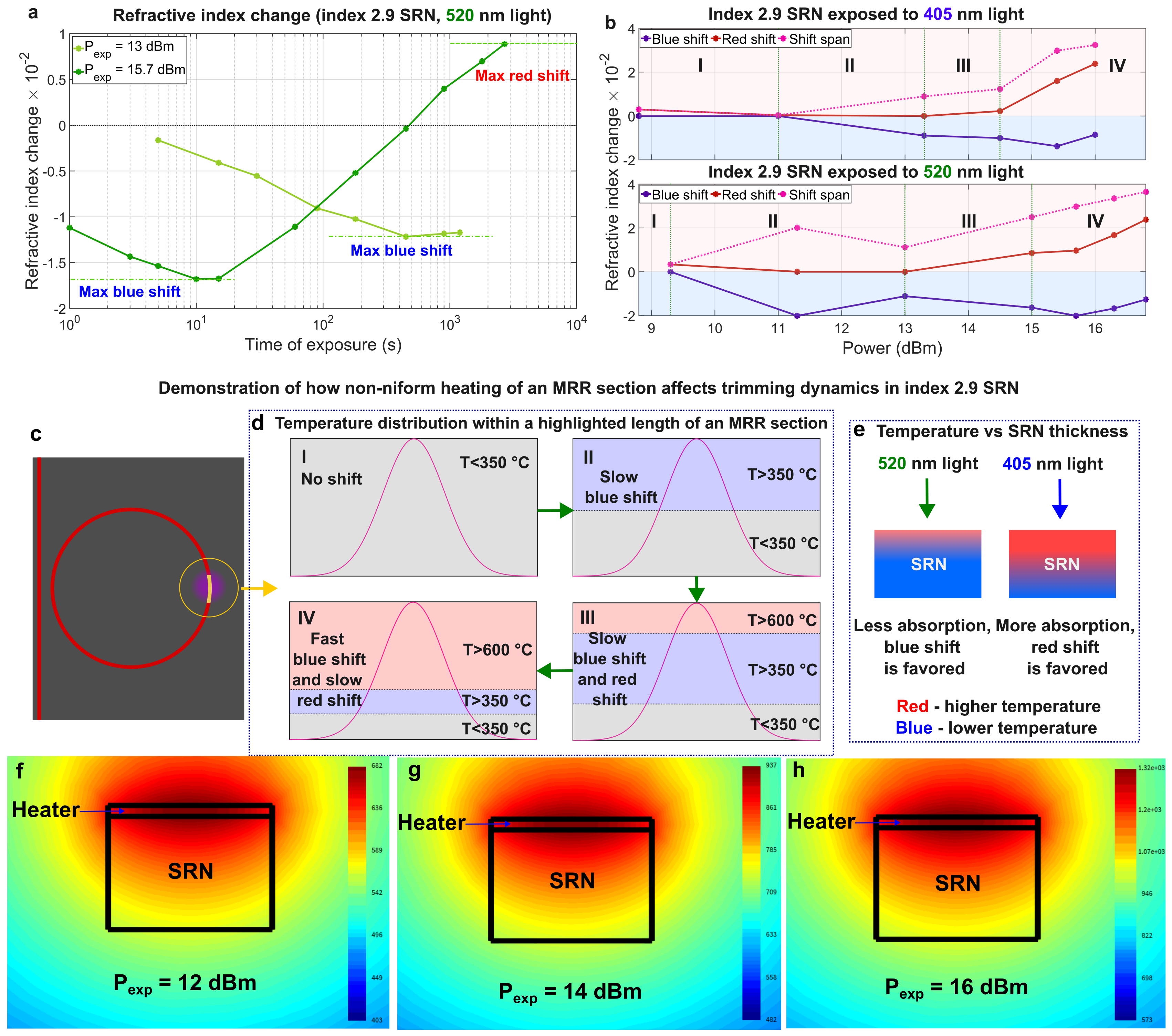}
\caption{(a) Dynamics of resonance shift in an SRN MRR when $n_{\text{srn}} = 2.9$ and $\lambda_{\text{exp}} = 520$ nm. (b) Refractive index change of SRN as a function of exposure power when $n_{\text{srn}} = 2.9$ (same as Figure 2c, presented here for convenience). (c) Image of an MRR showing how its section is exposed. (d) Demonstration of how the non-uniform distribution of heating within an exposed section of an MRR results in different trimming behavior (observed in Figure S4b). The grey, blue, and red areas represent no shift, blue shift, and red shift, respectively, due to annealing. The numbers I – IV correspond to cases described in Figure S4b. (e) Demonstration of how lower absorption of 520 nm light by an SRN waveguide results in larger blue shifts, while higher absorption of 405 nm light leads to red shifts, for the same level of power. Temperature distribution simulated in the index 2.9 SRN waveguide when exposed to 405 nm laser with the power of (f) 12 dBm, (g) 14 dBm, and (h) 16 dBm. Simulation was performed in Lumerical Device, the temperature is given in K. Thermal conductivity used in simulation is 1.7 W m-1 K-1, the thickness of the heater was 30 nm corresponding to the depth of the penetration of 405 nm light.
\label{figS4}}
\end{figure}

Figure S4b shows that the blue shift is more pronounced, even at lower exposure powers, when trimming is performed with 520 nm light, while the red shift is more prominent when using 405 nm light. This can be attributed to the higher absorption of 405 nm light by SRN, leading to greater heating and a stronger red shift, as illustrated in Figure S4e. Specifically, the extinction coefficients are 1.07 and 0.32 for 405 nm and 520 nm light, respectively, resulting in penetration depths of 30 nm and 130 nm, respectively. As light propagates through SRN, power attenuates according to the Beer–Lambert law, creating a temperature gradient. The lower absorption of 520 nm light results in a reduced temperature gradient, which favors blue shifts due to more uniform heating across the SRN waveguide. This uniformity makes it easier for sections of the ring to reach temperatures between 350°C and 600°C, where blue shifts are more likely. Figure S4e emphasizes this effect: with 520 nm light, only a small region at the top of the waveguide undergoes a red shift. However, with 405 nm light, stronger red shifts are seen at the top, and a steeper temperature gradient causes both red and blue shifts to partially offset, reducing the blue shift. In conclusion, 520 nm light is more favorable for bidirectional trimming, as it induces a strong blue shift due to the reduced temperature gradient, while a stronger red shift can be achieved by just increasing exposure power.
\\~\\
Here, we also examine the differences in trimming behavior between SRN with refractive indices of 2.4 and 2.9. Overall, during annealing, at lower temperatures (400–550 °C), the index 2.9 SRN undergoes a stronger blue shift (reduction in refractive index) compared to the index 2.4 SRN. This behavior is also reflected during laser trimming, where the index 2.9 SRN exhibits a more pronounced blue shift. 
\\~\\
The blue shift observed during annealing is attributed to the redistribution of bonds, such as Si-Si, Si-N, Si-H, and N-H [s1, s2]. Specifically, prior studies have reported an increase in Si-H and Si-N bond concentrations at the expense of N-H and Si-Si bond concentrations during annealing. We hypothesize that the higher concentration of Si-Si bonds in the index 2.9 SRN leads to a faster rate of bond redistribution, resulting in the formation of more Si-H and Si-N bonds, thereby reducing the refractive index. By contrast, the lower concentration of Si-Si bonds in the index 2.4 SRN limits the redistribution process, leading to a weaker blue shift. It is important to note that a higher concentration of Si-Si bonds in SRN correlates with a higher refractive index, as SRN becomes more silicon-like (closer to the refractive index of amorphous silicon, approximately 3.7) and less like stoichiometric Si\textsubscript{3}N\textsubscript{4} (refractive index 2.0).
\\~\\
At higher temperatures (\textgreater 600 °C), SRN experiences a refractive index increase due to thermal effects such as the dissociation of Si-H bonds, significant hydrogen desorption, reduced nitrogen content, and an increased concentration of Si-Si bonds [s1-s3]. The latter shifts SRN further toward a silicon-like composition, thereby increasing the refractive index. Notably, the index 2.9 SRN, which has a higher hydrogen content due to the increased use of SiH\textsubscript{4} precursor gas during deposition, is more affected by high-temperature annealing. As a result, its refractive index increases more significantly than that of the index 2.4 SRN under similar conditions.
\\~\\
Finally, the index 2.9 SRN exhibits a stronger bidirectional shift because it undergoes both a significant increase and decrease in refractive index. The spatial non-uniformity of the laser beam causes different regions of the rings to heat to varying temperatures, simultaneously inducing blue shifts (at T \textless 550 °C) and red shifts (at T \textgreater 600 °C). In contrast, the index 2.4 SRN primarily exhibits a red shift, as the blue shift is negligible in this case.
\\~\\
It is also evident that the blue shift occurs within just a few seconds at higher powers. The primary reason for this is that the blue shift happens at lower annealing temperatures (400-550°C), whereas the red shift occurs at higher temperatures (\textgreater 600°C). Consequently, the red shift requires higher powers than the blue shift. At lower power, the blue shift progresses more slowly (5-15 minutes), with negligible red shift. As a result, it takes longer for the blue shift to reach saturation, since there is no red shift to counteract or reverse it. As power increases, the red shift becomes evident, compensating for the blue shift and causing the trend to change from blue to red. In other words, at higher powers, the red shift dominates; it first slows the blue shift and eventually reverses it. Therefore, at higher powers, the time required for the red shift to reverse the blue shift is shorter, as red shift gets increasingly stronger.
\\~\\
Also, Figure S4f–S4h illustrates the simulated temperature distribution in Lumerical Device for SRN with a refractive index $n_{srn}$ = 2.9. In these simulations, we assumed a thermal conductivity of 1.7 W m\textsuperscript{-1} K\textsuperscript{-1}. It should be noted, however, that the exact thermal conductivity of SRN is not precisely known, as it depends on several factors, including the deposition recipe, hydrogen concentration, and thin-film thickness. Reported values for PECVD SRN thermal conductivity range widely from 0.45 to 4.5 W m\textsuperscript{-1} K\textsuperscript{-1} [s4]. The chosen value of 1.7 W m\textsuperscript{-1} K\textsuperscript{-1} is approximately in the middle of this range and provides simulation results that align well with the measured data shown in Figure S4b.
\\~\\
For example, at an exposure power of 12 dBm, the maximum simulated temperature is 682 K (409 °C), which is sufficient to induce a blue shift in specific sections of the waveguide. When the exposure power is increased to 14 dBm, the maximum temperature rises to 937 K (664 °C), a range where red shift begins to dominate in some sections of a waveguide, although blue shift effects are still dominant. Finally, at 16 dBm, the peak temperature reaches 1320 K (1046 °C), which would lead to a pronounced red shift due to high-temperature annealing effects.
\\~\\
By comparing these simulated results with the experimental data in Figure S4b, we observe a reasonable agreement. This suggests that the assumed thermal conductivity value is a practical approximation for the SRN material used in our study. We conclude that the relatively low thermal conductivity of SRN (compared to that of c-Si, which is 148 W m\textsuperscript{-1} K\textsuperscript{-1} [s5]) enables the generation of high local temperatures during laser exposure, facilitating effective bidirectional trimming of the waveguide refractive index.

\section{Minimizing scattering loss during trimming}

\begin{figure}[ht]
\includegraphics[width=1\linewidth]{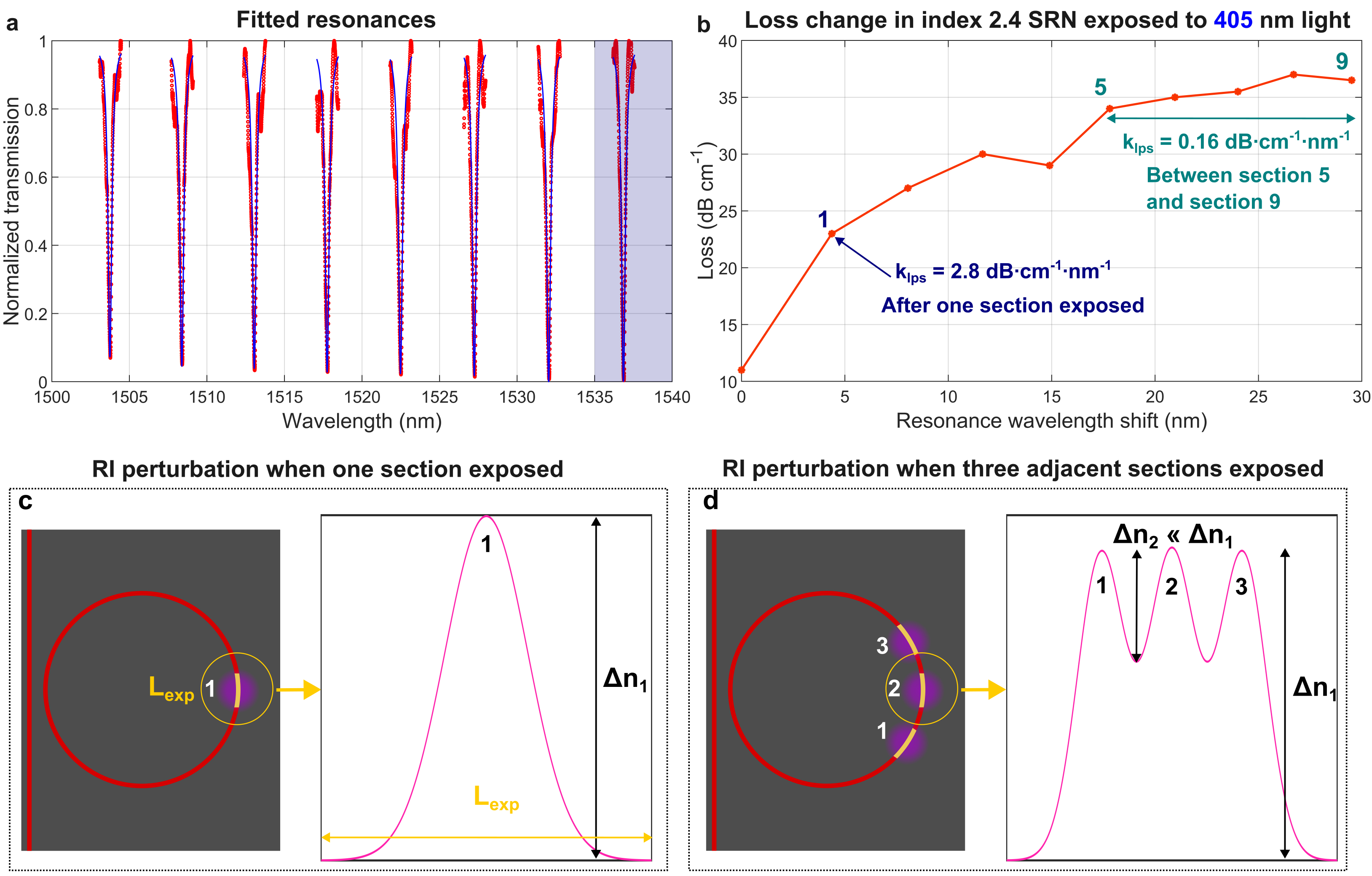}
\caption{(a) An example of transmission spectrum with fitted resonances of an exposed MRR resonator when $n_{\text{srn}} = 2.4$, $\lambda_{\text{exp}} = 405$ nm. The blue area denotes the wavelength range where losses were measured. Blue curves represent a fitting function, red circles – measured data. (b) Loss change as a function of shifted resonance wavelength after trimming of an MRR resonator when $n_{\text{srn}} = 2.4$, $\lambda_{\text{exp}} = 405$ nm (measured in range 1535 nm – 1540 nm). Here numbers mean how many sections of the same MRR were exposed. The parameter $k_{\text{lps}}$ is loss per wavelength shift, it determines how much loss increases (in dB cm$^{-1}$) per 1 nm resonance shift. (c) Illustration of how exposure of one section of an MRR results in RI perturbation $\Delta n_1$ responsible for scattering losses. (d) Illustration of how exposure of three adjacent sections of an MRR results in RI perturbation $\Delta n_1$ and $\Delta n_2$ responsible for scattering losses.
\label{figS5}}
\end{figure}

In this section, we examine how different trimming approaches impact scattering losses. Figure S5a illustrates the process of determining optical losses by fitting MRR resonances. For consistency, all loss measurements were conducted in the wavelength range of 1535 nm – 1540 nm. Figure S5b shows the increase in scattering losses as a function of the resonance wavelength shift induced by 405 nm light trimming when $n_{\text{srn}} = 2.4$. Each point in this plot represents a different adjacent section of the same MRR exposed to 405 nm light, with a spacing of 17 $\mu$m between sections.
\\~\\
Exposure was organized such that the first section (labeled as section 1 in Figure S5b) was exposed for 45 minutes, the second for 30 minutes, and the remaining sections for 15 minutes each. It is evident that exposing the first section for 45 minutes led to a considerable increase in scattering losses, as expected: the estimated refractive index (RI) increase was approximately 0.1, changing from 2.4 to 2.5. However, this increase in losses diminished significantly with exposure of the adjacent sections. Indeed, when only the first section was exposed for 45 minutes, the loss increase per wavelength shift, $k_{\text{lps}}$, was 2.8 dB cm$^{-1}$ nm$^{-1}$. Notably, this value dropped to 0.16 dB cm$^{-1}$ nm$^{-1}$ between sections 5 and 9, indicating that a 1 nm shift would lead to an average increase of only 0.16 dB cm$^{-1}$, which is acceptable for many applications where expected resonance shifts remain below the FSR of an MRR (typically 1 nm – 5 nm). We propose that $k_{\text{lps}}$ can be further minimized by exposing the MRR uniformly rather than in discrete steps.
\\~\\
Figure S5c and S5d clarify how a more uniform exposure reduces scattering losses by decreasing RI perturbations. In this example, closely spaced exposed sections lead to a more uniform RI increase ($\Delta n_2 \ll \Delta n_1$), resulting in less scattering. Thus, ensuring uniform exposure of the MRR is essential to mitigate scattering losses, which can otherwise be significantly detrimental, as demonstrated in this section.

\section{Negligible contribution to trimming from PECVD SiO$_2$ and remaining HSQ}

In this section, we investigate whether PECVD SiO2 or remaining HSQ can account for the permanent resonance shifts observed in SRN MRRs. To explore this, we fabricated c-Si MRRs (waveguide width: 450 nm, waveguide thickness: 220 nm), as shown in Figure S6a, and exposed one section of the MRR to 520 nm light (Figure S6b). The fabrication process followed the same flow detailed in the Methods and materials section, with two exceptions: the Si device layer was not removed, and SRN was not deposited. The primary objective was to assess whether permanent resonance shifting occurs in the absence of SRN.

\begin{figure}[ht]
\includegraphics[width=1\linewidth]{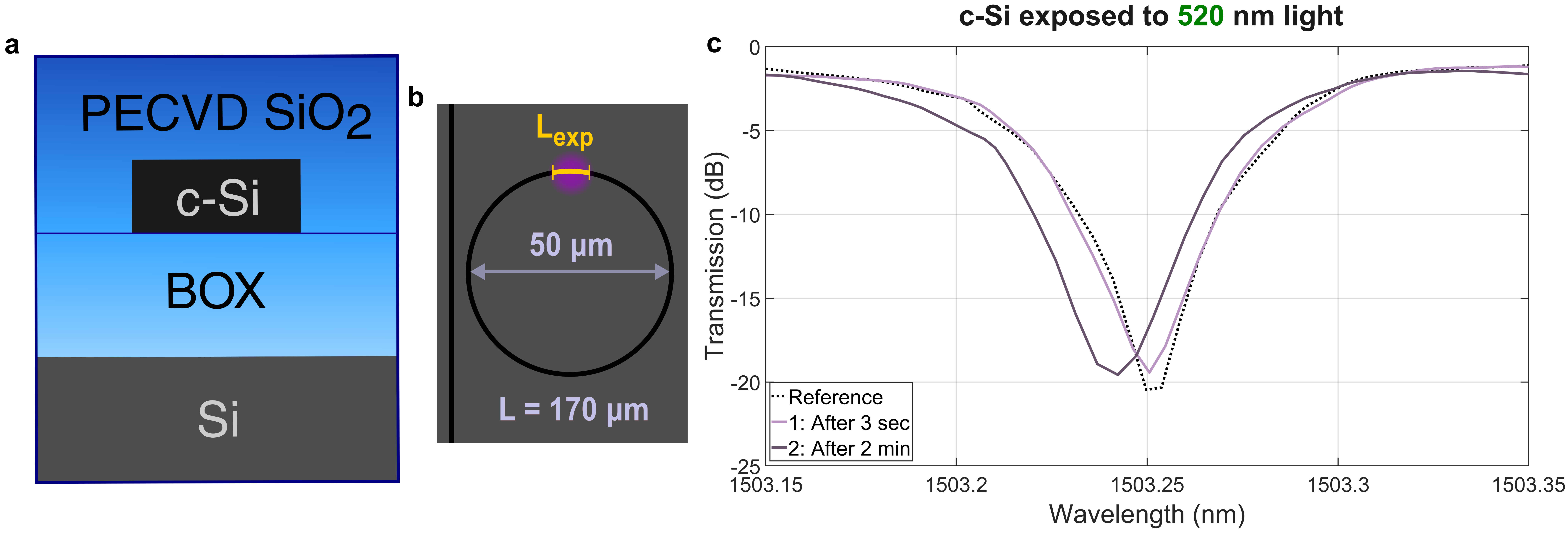}
\caption{(a) Cross-section of a c-Si waveguide used in experiments. Here, BOX refers to buried oxide. (b) Image of an MRR showing how its section is exposed (violet circle). The MRR diameter is 50 $\mu$m, and the ring length is 170 $\mu$m. (c) Transmission spectra of the probing signal in c-Si MRRs after exposure to 520 nm light, $P_{\text{exp}} = 16.7$ dBm. The dotted black curve indicates the initial position of resonance (reference). The labels describe exposure time.
\label{figS6}}
\end{figure}

Based on the transmission spectra (Figure S6c), even after 2 minutes of exposure, the impact on the resonance position is negligible. Specifically, the observed shift of -10 pm is far too small to account for the substantial shifts seen in SRN MRRs. Thus, we conclude that neither PECVD SiO$_2$ nor residual HSQ contribute to the resonance shifts observed in SRN, regardless of the refractive index.

\medskip
\textbf{References for Supplementary Information} \par

[s1] S.~S.~He, V.~L.~Shannon,
\newblock \emph{Proc. Int. Conf. Solid-State IC Technol.} \textbf{1995}, 269.

[s2] F.~L.~Martínez, A.del Prado, I.~Martil,
\newblock \emph{Phys. Rev. B} \textbf{2001}, \emph{63}, 245320.

[s3] P.~Dani, M.~Tuchen, B.~E.~Meli, J.~Franz,
\newblock \emph{Micro Nano Eng.} \textbf{2024}, \emph{100}, 100291.

[s4] M.~Hadi, S.~Pailhès, R.~Debord, A.~Benamrouche, E.~Drouard, T.~Gehin, C.~Botella, J.-L.~Leclercq, P.~Noe, F.~Fillot, V.~M.~Giordano,
\newblock \emph{Materialia} \textbf{2022}, \emph{26}, 101574.

[s5] M.~Jacques, A.~Samani, E.~El-Fiky, D.~Patel, Z.~Xing, D.~V.~Plant,
\newblock \emph{Opt. Express} \textbf{2019}, \emph{27}, 10456.

\end{document}